\documentclass{aa}  

\usepackage{graphicx}
\usepackage{txfonts}
\usepackage{amsmath}	
\usepackage{amssymb}	
\usepackage{xcolor}
\usepackage{ulem} 
\usepackage{caption}
\usepackage{placeins}
\usepackage[colorlinks=true,citecolor=blue,linkcolor=blue]{hyperref}
\newcommand{\dd}{\mathrm d}

\newcommand{\ob}{Cygnus~OB2\xspace}
\newcommand{\fmlat}{\textit{Fermi}-LAT\xspace}

\newcommand{\kms}{\,\mathrm{km}\,\mathrm{s}^{-1}}
\newcommand{\Msyr}{\,\mathrm{M}_\odot\,\mathrm{yr}^{-1}}
\newcommand{\ergs}{\,\mathrm{erg}\,\mathrm{s}^{-1}}

\newcommand{\eVccm}{\,\mathrm{eV}\,\mathrm{cm}^{-3}}
\newcommand{\pccm}{\,\mathrm{cm}^{-3}}
\newcommand{\Msun}{\,\mathrm{M}_\odot}

\newcommand{\ur}[1]{\mathrm{#1}}

\begin{document} 
\makeatletter
\renewcommand*\aa@pageof{, page \thepage{} of \pageref*{LastPage}}
\makeatother

\title{Deciphering the gamma-ray emission in the Cygnus region}
\titlerunning{Deciphering the $\gamma$-ray emission in the Cygnus region}
\authorrunning{L. H\"arer et al.}

\author{L. Härer \inst{1}\fnmsep\thanks{\email{lucia.haerer@mpi-hd.mpg.de}}, 
        T. Vieu \inst{1}\fnmsep\thanks{\email{thibault@mpi-hd.mpg.de}},  F. Schulze\inst{1}, C. J. K. Larkin\inst{1,2,3}, and B. Reville\inst{1}}

\institute{
Max-Planck-Institut für Kernphysik, Saupfercheckweg 1, 69117 Heidelberg, Germany \and
Zentrum f\"{u}r Astronomie der Universit\"{a}t Heidelberg, Astronomisches Rechen-Institut, M\"{o}nchhofstr. 12-14, 69120 Heidelberg, Germany \and
Max-Planck-Institut f\"{u}r Astronomie, K\"{o}nigstuhl 17, D-69117 Heidelberg, Germany}

\date{Received May 15, 2025; accepted August 29, 2025}

 
\abstract
{The Cygnus region is a vast star-forming complex harbouring a population of powerful objects, including massive star clusters and associations, Wolf-Rayet stars, pulsars, and supernova remnants. The multi-wavelength picture is far from understood, in particular the recent LHAASO detection of multi-degree scale diffuse $\gamma$-ray emission up to PeV energies.
We aim to model the broadband $\gamma$-ray data, discriminating plausible scenarios amongst all candidate accelerators. We consider in particular relic hadronic emission from a supernova remnant expanding in a low-density environment and inverse Compton emission from stellar-wind termination shocks in the Cygnus~OB2 stellar association. 
We first estimate the maximum particle energy from a 3D hydrodynamical simulation of the supernova remnant scenario.
The transport equation is then solved numerically to determine the radial distribution of non-thermal protons and electrons.
In order to compute synthetic $\gamma$-ray spectra and emission maps, we develop a 3D model of the gas distribution. This includes, firstly, a HI component with a low-density superbubble around Cygnus~OB2 and, secondly, molecular clouds lying at the edge of the superbubble and in the foreground.
We find that a powerful, ${\sim}50$\,kyr-old supernova remnant can account for both the morphology and spectrum from $10$\,TeV--PeV. 
At PeV energies, the microquasar Cygnus X-3 and diffuse galactic cosmic rays might also contribute to the flux.
Below about 10\,TeV, hadronic models are incompatible with the expected existence of a superbubble centred on Cygnus~OB2. Instead, the spectrum is well fitted with inverse Compton emission from electrons accelerated at stellar-wind termination shocks in Cygnus~OB2 in line with existing multi-wavelength limits.
{}
    }


\maketitle

%

\section{Introduction}
\label{sec:intro}
The Cygnus region is the most massive star-forming complex in the local Arm. It displays rich multi-wavelength emission features from radio to ultra-high energy (${>}100$\,TeV) $\gamma$-rays, revealing a highly structured environment shaped by the feedback of numerous stellar objects. Unfortunately, the region lies in a high extinction patch of the sky, hidden behind a dark lane of gas and dust called the Cygnus Rift \citep[see][for a review]{Reipurth2008}. At a galactic longitude of about 80$^\circ$, it is also seen tangentially to the direction of galactic rotation, which complicates distance estimations, making it difficult to disentangle superimposed features along the line-of-sight. Despite decades of extensive studies, the observations remain poorly understood, in particular the nature and origin of the structures visible on large scales.

Cygnus was first detected as a diffuse radio source by \citet{Piddington1952}, who named it Cygnus-X (``X''  for ``extended''). On top of the diffuse emission, \citet{Downes1966} identified 20 individual HII regions, later assigned DR4--DR23, which appear as prominent radio hotspots. The only source of non-thermal (synchrotron) radio emission is the supernova remnant $\gamma$-Cygni \citep{xu2013radio}. The diffuse thermal component comes from material ionised by massive stars, in particular the stellar association \object{Cygnus~OB2}, which is thought to be responsible for the diffuse emission in a Strömgren sphere of 2$^\circ$ in radius. In addition to the diffuse emission, the compact radio hotspots, coincident with dense clumps, originate from massive stars and protostars within the surrounding clusters and star-forming clouds \citep{Knodlseder2002,Knodlseder2004}. 
Enclosing the diffuse radio emission is a large ($18^\circ \times13^\circ$) elliptic ring of soft X-rays \citep[0.5--2\,keV,][]{Cash1980}. This X-ray shell, historically referred to as the ``Cygnus Superbubble'', is thought to have been created by the collective feedback of massive stars and past supernovae in Cygnus~X over the past tens of millions of years \citep{kimura2013,Bluem2020}, though the possibility of a chance coincidence of X-ray emission along the line-of-sight is not excluded \citep{Uyaniker2001}.
Another hint of large-scale feedback in Cygnus is the presence of a ring of H$\alpha$ filaments \citep{Dickel1969} coincident with the X-ray shell. These observations are consistent with the existence of an ionised low-density cavity of radius 100--150\,pc roughly centred on \ob, which could act as the source of ionising radiation and mechanical power. This notion is consistent with the current stellar-wind power in Cygnus~OB2 ($2\mbox{--}3 \times 10^{38}\ergs$, \citealt{Vieu2024CygnusSimu}) integrated over several Myr of stellar activity. Detection of diffuse X-ray emission in the close vicinity of \ob at 1--3\,keV suggests that stellar winds are indeed interacting close to the centre of the association \citep{AlbaceteColombo2023}. Despite its relevance, the relation between the small-scale stellar feedback and the large-scale features is still not settled, and has received scarce attention in the recent literature.

Stellar feedback in such a complex environment is not expected to produce spherically symmetric bubbles, but rather intricate, asymmetric structures. Critical factors include the wind properties of the most massive stars, the inhomogeneous interstellar medium, as well as the past activity of supernovae and older star clusters in the region (e.g. NGC6913 and NGC6910, see \citealt{Almeida2023}). Mid-infrared observations of Cygnus~X reveal a rich morphology in the warm dust emission with bright photodissociation regions, molecular ridges, extended pillars, filaments, and parsec-scale globules \citep{Schneider2006}. Most likely, these structures were heated and shaped by the \ob association. Far-infrared imaging \citep{schneider16} shows a complex web of colder molecular clouds interacting with multiple HII regions including nascent star clusters close to \ob, which hints at triggered star-formation events. These infrared observations correlate well with the CO intensity, which traces molecular hydrogen. The molecular complex is usually divided between the northern ($l>80^\circ$) and the southern ($l<80^\circ$) clouds. In Cygnus North lies the well-studied dense star-forming molecular ridge DR21 \citep[e.g.][]{Hennemann2012_DR21,Bonne2023_DR21}. In Cygnus South, the CO emission is more diffuse yet tends to coincide with bright mid-infrared emission, in particular at the location of the molecular cloud L889, and radio hotspots (e.g. DR9, DR12, DR13, and DR15, see \citealt{Schneider2006}). It is possible that the molecular material was shaped by interacting superbubble cavities blown by \ob, NGC~6913 (or the more extended Cygnus OB9 association) and NGC~6910 \citep[e.g.][]{Schneider2007}, although the difficulty in measuring accurate distances to both cloud material and star clusters complicates any such interpretation. Recent progress has been made by \citet{Zhang2024} who applied an extinction jump method to high resolution CO data, concluding that the molecular gas is organised in multiple layers between 0.8 and 2\,kpc from the Sun. In particular, the method successfully disentangles the foreground Rift component, which appears to extend out to approximately 1.1\,kpc and seems to contain more gas than previously anticipated in earlier work by \citet{Schneider2006}. An intermediate layer at 1.3\,kpc, with counterparts both in Cygnus North \citep[W75N and DR17, see also][]{Gottschalk2012} and Cygnus South, also appears in the foreground of the active star-forming complex, the latter being located around 1.6--1.8\,kpc. Unfortunately the extinction is too high in Cygnus to track extinction jumps out to 2\,kpc. It is not presently possible to precisely disentangle the foreground clouds from the molecular web interacting with the star-forming complex at larger distances.

Finally, identifying coherent stellar groups is not straightforward. The stellar population is scattered and diverse, with a large age-spread and possible past supernova events \citep[e.g.][]{Reipurth2008,Comeron2012,Comeron2020,Berlanas2020}. Modern clustering algorithms do not identify Cygnus OB1, OB8 and OB9 as genuine associations \citep{Quintana2021}, and {\textit{Gaia}} measurements indicate that even Cygnus~OB2, the largest and most powerful group of stars, is not a single coherent group but a superposition of several groups along the line-of-sight \citep{Berlanas2019,Orellana2021}, the most massive being located at about 1.6--1.7\,kpc. The latter is what we refer to as ``Cygnus~OB2'' in the remainder of the present work (see Section~\ref{sec:ob2}).

\medbreak

It might be unsurprising that such a complex region with a long history of star-forming activity is a bright source in the $\gamma$-ray sky. Extended GeV sources in Cygnus were first detected by \citet{Cygnus_gamma_EGRET} and later associated with the $\gamma$-Cygni supernova remnant \citep{Brazier1996} and the Dragonfly pulsar wind nebula \citep[around PSR J2021+3651,][]{Roberts2002}. An extended TeV source coincident with \ob was discovered by \citet{Gammaobs_HEGRA_2002} and \citet{Abdo2007} and later associated with the pulsar wind nebula powered by PSR J2032+4127 \citep{Camilo2009}.

On top of these three identified $\gamma$-ray sources, the \fmlat collaboration discovered an extended diffuse $\gamma$-ray source coincident with the radio emission in Cygnus~X, filling a region delimited by bright infrared photodissociation regions \citep{Cygnus_FERMI}. They demonstrated that in-situ particle acceleration was necessary to account for the observation. This breakthrough revived the high-energy community's interest in star-forming regions as cosmic-ray accelerators, and was followed up with detections by \citet{dataArgo2014}, \citet{gamma_Cygnus2018_VERITAS}, \citet{Tibet2021}, \citet{gammaCygnus_HAWC2021} and \citet{Lhaaso2024}. The centroid of the diffuse emission lies at the heart of Cygnus~X, approximately coincident with the position of Cygnus~OB2 below 1\,TeV, and loosely correlated with the northern and southern molecular clouds at ultra-high energy. The spectrum is hard in the GeV band ($\ur{d}N/\ur{d}E \sim E^{-2}$) but turns over around 1\,TeV ($\ur{d}N/\ur{d}E \sim E^{-3}$). The steep component measured by \citet{Lhaaso2024} extends up to 2\,PeV with a flux substantially above the estimate for the galactic diffuse background, which suggests the presence of an extreme particle accelerator in the region. This accelerator is unlikely related to PSR J2032+4127 or $\gamma$-Cygni at their current stages of evolution, as the spectra of these sources cut off at energies ${\lesssim}100$\,TeV and their integrated flux is subdominant with respect to the extended diffuse component \citep{Fleischhack2019,gammaCygnus_HAWC2021,MAGICgCYGNI2023,Lhaaso2024}.

The nature of candidate PeV accelerators in Cygnus has been debated in recent years. \citet{Bykov2022Cygnus} proposed a scenario of particle acceleration in supersonic turbulence excited by the wind feedback in Cygnus~OB2. They model a cavity of radius 55\,pc filled with a turbulent plasma of average velocity $1500\kms$, though this requires an energy budget that cannot be supplied by the massive stars in Cygnus~OB2 alone for any reasonable plasma density. Such a scenario also fails to account for the ultra-high energy detection by LHAASO, because of fundamental limitations of second-order acceleration mechanisms \citep{vieu2022Emax,Vieu2024core}. \citet{Menchiari2024} postulated the existence of a large-scale wind termination shock around Cygnus~OB2 and applied a particle acceleration and transport model in spherical symmetry \citep{morlino2021}, showing good agreement with \fmlat and HAWC data. This mechanism is however unlikely to produce multi-PeV protons (see Sect.~\ref{sec:candidate_accelerators}), and it has further been shown that Cygnus~OB2, as a loose association, is not compact enough to produce a large-scale cluster wind \citep{Vieu2024CygnusSimu}. In addition, hadronic models for $\gamma$-ray emission in star-forming environments generally suffer from the fundamental caveat that they simultaneously need i) a low-density environment which allows the persistence of strong shocks and inside which particles are confined and ii) a high density environment for efficient proton-proton interactions. This is why leptonic scenarios were previously favoured in star-forming regions \citep[see][for the case of Westerlund 1]{Haerer2023}. A leptonic origin for the $\gamma$-ray emission in Cygnus was excluded by \citet{gammaCygnus_HAWC2021} arguing that in a magnetic field of 20\,\textmu G and a density of 30\,cm$^{-3}$ one would expect a detection of synchrotron and bremsstrahlung radiation beyond the limits set by radio and X-ray observations \citep[see also][]{Guevel2023}. These values for the magnetic field and gas density are however far in excess of what one would expect in a region dominated by strong stellar feedback. Magnetohydrodynamic simulations of star clusters show that the magnetic field drops rapidly in the cluster wind and is no more than a few \textmu G in the wind-blown superbubble \citep{Haerer2025}. Hydrodynamic simulations, along with analytic theory \citep{weaver1977} predict a density of at most a few 0.1\,cm$^{-3}$ within a few tens of parsecs from Cygnus~OB2. The values quoted by the HAWC collaboration are taken from \citet{Cygnus_FERMI}, where they were used only to estimate the properties of the Cygnus molecular clouds, not those of the ionised gas around Cygnus~OB2. Unfortunately, the properties of this gas remain unknown -- though there is a strong theoretical argument in favour of a low density cavity excavated by winds from massive stars in \ob \citep[e.g.][]{Bluem2020,Vieu2024CygnusSimu}, in addition to the observational hints outlined above. 
This promotes consideration for a leptonic origin of the diffuse $\gamma$-ray emission, in particular given the strong photon field in the vicinity of Cygnus~OB2 \citep[see][and Sect.~\ref{sec:phfields}]{schneider16}. Such models \citep[e.g.][]{Banik2022,Astiasarain2023} cannot however account for the ultra-high energy component of the spectrum, since the cooling time of PeV electrons is much shorter than the acceleration time at a non-relativistic shock under usual circumstances. 

The aim of the present work is to clarify the origin of the broadband $\gamma$-ray emission in Cygnus, in light of the current multi-wavelength picture. In Sect.~\ref{sec:candidate_accelerators} we start by reviewing the stellar objects observed in the region and conclude that all are disfavoured as the source of the non-thermal PeV radiation at their current stages of evolution. In Sect.~\ref{sec:SNRsimulation} we make the case for a past supernova event in Cygnus~OB2 and explore this scenario by performing a detailed hydrodynamic simulation of a supernova remnant in a Cygnus-like stellar association. Section~\ref{sec:gas} is devoted to a detailed modelling of the gas, disentangling the foreground components. We then proceed to build a lepto-hadronic model to show how the diffuse $\gamma$-ray emission can be produced by the joint action of stellar winds and a past supernova. Section~\ref{sec:modelling} introduces the methods. The supernova remnant model is then discussed in Sect.~\ref{sec:gamma-sn} and the wind model in Sect.~\ref{sec:gamma-ob2}. We present our conclusions in Sect.~\ref{sec:conclusion}.

\section{Candidate particle accelerators in Cygnus}
\label{sec:candidate_accelerators}
The Cygnus region hosts a diverse population of powerful objects, including massive star clusters and associations \citep[e.g.][]{Comeron2020,Quintana2021}, Wolf-Rayet stars, pulsars \citep[e.g.][]{Lyne2015,Jin2023}, supernova remnants \citep[e.g.][]{Leahy2013} and microquasars \citep[e.g.][]{Aleksic2010}. In the following section we review the 
extreme objects in Cygnus and discuss possible acceleration mechanisms.

\subsection{Cygnus~OB2}
\label{sec:ob2}
The most notable stellar group in Cygnus is the \ob association, which is by far the most massive and lies roughly at the centre of the region of diffuse radio emission (``Cygnus X''). Other OB associations \citep{Quintana2021} or young star clusters (e.g. Berkeley 86, NGC6913, NGC6910) are substantially less massive and most likely older according to estimates by \citet{Cantat2020}. Hosting a rich population of more than 70 O stars and three Wolf-Rayet stars, \ob has a mechanical wind luminosity of $2\mbox{--}3\times 10^{38}\ergs$ (see \citealt{Vieu2024CygnusSimu} for a review of the main properties of the association). It is worth noting that \ob is not a compact star cluster but a rather loose OB association, with a half-mass radius of about 15\,pc \citep{wright15}. As such, the stellar population is heterogeneous and likely not coeval, with star ages ranging between 3 and 7\,Myr and hints of multiple star-forming episodes \citep[e.g.][]{Berlanas2020}. Part of the heterogeneity is due to the superposition of several stellar groups along the line-of-sight at about 1.3, 1.5 and 1.7\,kpc \citep{Orellana2021}. This complication can also explain some earlier confusion in distance estimates (see Appendix~\ref{appendix:distancetoOB2}). In light of {\textit{Gaia}}~DR2, it seems that the main group is located at about 1.6--1.7\,kpc, the uncertainty being mostly due to the {\textit{Gaia}} zero-point offset \citep[e.g.][]{Lindegren2018}. We stress that the distance of 1.4\,kpc inferred by \citet{Rygl2012} employing maser measurements, which is frequently quoted in the high-energy community, is no longer believed to be the distance to \ob, but rather the distance to the molecular complex in Cygnus North, which lies in the foreground of the association. 

\subsection{Other objects}
\begin{figure}
        \centering
        \includegraphics[width=\linewidth]{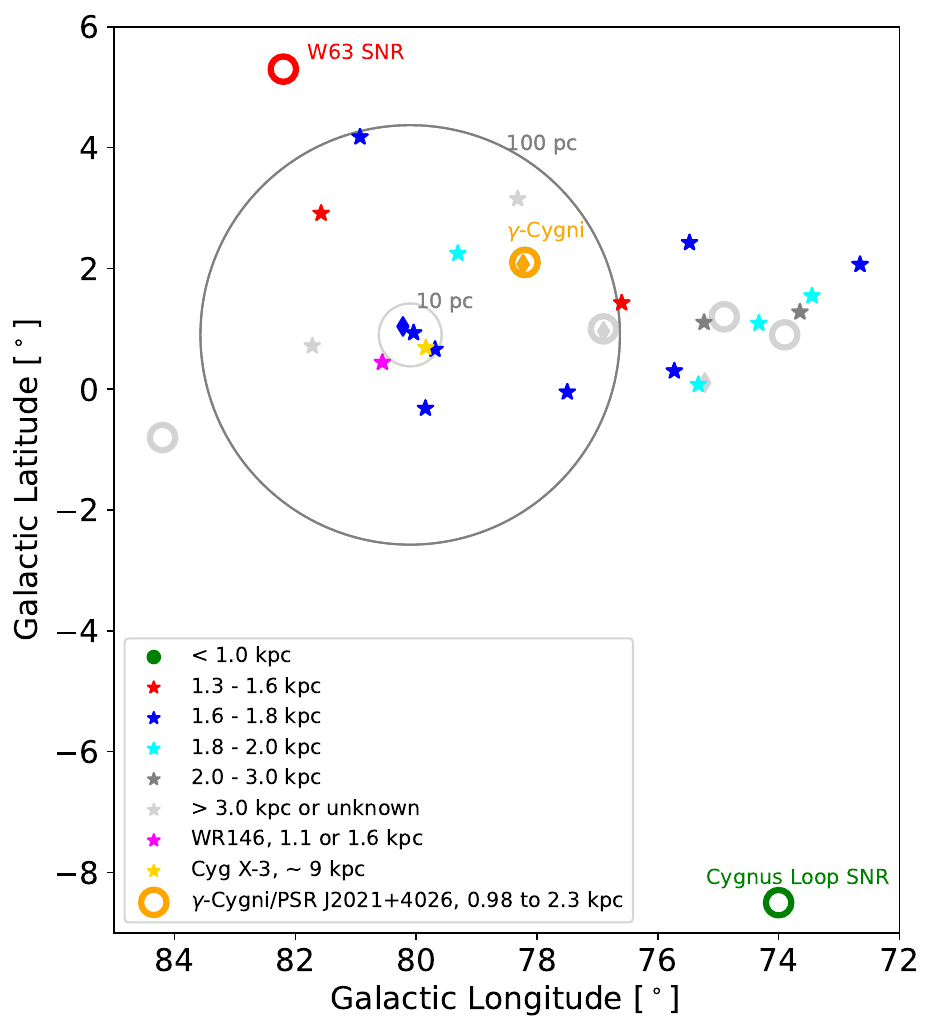}
        \caption{Powerful objects in the Cygnus region. Stars: Wolf-Rayet stars from \citet[][\url{http://pacrowther.staff.shef.ac.uk/WRcat/index.php}]{Rosslowe2015}. Diamonds: pulsars. The distance to  PSR J2032+4127 has been revised to 1.76~kpc by association with its Be star companion MT91 213. Rings: location of supernova remnants from \citet[][\url{https://www.mrao.cam.ac.uk/surveys/snrs}]{Green2024_SNRcatalogue,Green2025_SNRcatalogue}. Different methods give distances to $\gamma$-Cygni between 0.98 to 2.3\,kpc. The circles labelled ``10\,pc'' and ``100\,pc'' are centred on Cygnus~OB2, assuming a distance of 1.65~kpc to the association. }
        \label{fig:candidate_accelerators}
    \end{figure}

A map showing the location of the powerful objects in the Cygnus region is provided in Fig.~\ref{fig:candidate_accelerators}. 
Tens of Wolf-Rayet stars are scattered throughout the region, either clustered or isolated \citep{Rosslowe2015}. The most powerful are WR144 (most likely a member of \ob), WR142 (likely a member of Berkeley~87) and WR147, each one having a mechanical power of about $10^{38}\ergs$ \citep{sander2019,hamann2019}. In addition, over 20 red supergiants are observed \citep{Comeron2020}. Three pulsars have been detected in the region: PSR J2021+3651 (spin-down power ${\sim}3 \times 10^{36}\ergs$), which powers the Dragonfly pulsar wind nebula, PSR J2021+4026 (spin-down power ${\sim}10^{35}\ergs$), 
and finally PSR J2032+4127 (spin-down power ${\sim}10^{35}\ergs$), which likely lies within the \ob association and suggests past supernova activity at the heart of the star-forming complex. However, only two supernova remnants are observed in the 1--3\,kpc range in the direction of Cygnus: the bright source $\gamma$-Cygni and the supernova remnant W63.

In the following we address whether any of these objects could potentially accelerate particles beyond 1\,PeV, as required to explain the recent LHAASO detection of ultra-high energy (${>}100\,$TeV) photons from the region.
A contribution from PSR J2021+3651 is excluded as the Dragonfly nebula is masked from the LHAASO analysis. A source coincident with PSR J2032+4127 in \ob is detected as a subdominant component with a low-energy (30 TeV) cut-off and is therefore unlikely to significantly contribute to the integrated diffuse $\gamma$-ray emission, although it might contaminate the significance map around \ob.

Of the three supernova remnants observed within 2~kpc toward the Cygnus region, W63 and the Cygnus Loop are older remnants that, while perhaps efficient accelerators in the distant past, are both too far away from the centre of emission to account for the integrated flux. $\gamma$-Cygni is a younger remnant \citep[about 5\,kyr,][]{Lozinskaya2000} which is bright in radio, X-rays, and $\gamma$-rays. It expands in a low-density environment (about 0.1\,cm$^{-3}$) to a present size of about 15\,pc with a shock velocity of nearly $1000\kms$ \citep{Lozinskaya2000}. 
This low-density environment could be part of the superbubble blown by Cygnus~OB2 or the cavity from NGC~6910 (depending on its position along the line-of-sight). The $\gamma$-ray spectrum associated with $\gamma$-Cygni extends to tens of TeV \citep{Fleischhack2019} and is thought to originate from proton-proton interactions with a nearby cloud \citep{MAGICgCYGNI2023}. This component is already subtracted in the data analysis of \fmlat, HAWC, and LHAASO. It is however not excluded that ultra-high energy $\gamma$-ray emitting particles were accelerated in the early expansion phase of the remnant and already escaped the $\gamma$-Cygni region to fill a network of cavities in the Cygnus star-forming complex, eventually reaching the molecular clouds on its edge. 

Finally, the microquasar Cygnus X-3 is seen in coincidence with Cygnus~OB2 and therefore lies in the centre of the diffuse $\gamma$-ray emission. Similar microquasars have been recently recognised as efficient particle accelerators up to 100s of TeV \citep{LHAASO2024_microquasar}. At an estimated distance of 9\,kpc, Cygnus X-3 is unlikely to contribute to the bulk of the emission. Particles accelerated in the system would have to be confined in a physical radius of over 300\,pc to reproduce the 2$^\circ$ extent of the diffuse emission. However, a subdominant contribution at ultra-high energies
appears plausible, as we discuss in Sect.~\ref{sec:gamma-sn}.
Given these details, the best accelerator candidates in the Cygnus region are most likely linked to the energy input from massive stars, including the possibility of a supernova remnant in the low-density cavity around the association.

\subsection{Particle acceleration}
Winds from massive stars or star clusters are thought only able to accelerate PeV particles if they are exceptionally powerful and strongly magnetised. The physical requirements are detailed in \citet{vieu2022Emax} and  \citet{Haerer2025}. A key limitation in wind acceleration models comes from the necessity to overcome adiabatic losses when accelerated particles propagate in the radially expanding region upstream of the wind termination shock. Note that the argument applies both in the case of an isolated massive star or a compact star cluster powering a collective outflow. Disregarding the downstream residence time, which might be shortened by field amplification, shock induced turbulence or wind-wind collisions, the acceleration rate at a strong shock reads $\dot{p}_\ur{acc} \approx V_{\rm w}^2 p/(4 D)$, where $V_{\rm w}$ is the wind speed and $D$ is the upstream diffusion coefficient \citep{lagage1983, drury1983}. This must always be larger than the adiabatic loss rate in the upstream, which reads $\dot{p}_\ur{loss} \geq 2 V_\ur{w} p/(3R_{\rm WTS})$, where $R_{\rm WTS}$ is the radius of the wind termination shock. The particle energy is therefore bounded as $D(E) < 3 R_{\rm WTS} V_{\rm w}/8$. The most efficient acceleration is reached when the diffusion is in the Bohm limit in a magnetic field of strength $B$: $D(E) = E c/(\xi q B)$ with $q$ the charge of the particle, $c$ the speed of light and $\xi \leq 3$. The maximum energy is then limited as $E_{\rm max} < q B R_{\rm WTS} V_{\rm w}/c$, which is recognised as the Hillas limit in the upstream region \citep{hillas1984}.

Independently of the assumption made for the generation or amplification of the magnetic field, the magnetic field strength is limited by the requirement that the wind must be super-Alfv\'enic for acceleration to proceed. This sets a loose limit on $B$ as $B = \sqrt{4 \pi \rho} V_{\rm w}/ M_{\rm A}$, where $\rho$ is the gas density and efficient particle acceleration requires $M_{\rm A}$ substantially exceeds unity. 
Conservation of mass in a spherical wind satisfies $\rho = \dot{M}/(4 \pi  R_{\rm WTS}^2 V_\ur{w})$, which leads to $E_{\rm max} \ll q \sqrt{ V_{\rm w} L_{\rm w}}/c$, where $L_{\rm w}=0.5\dot{M}V_\ur{w}^2$ is the mechanical power and $\dot{M}$ the mass-loss rate of the central source. Remarkably, this upper limit only depends on the properties of the star and is in particular independent of the geometry of the shock.
The limit barely reaches 1\,PeV for the most powerful stars in the heart of Cygnus~OB2 (WR144, WR146, Schulte~7). In practice, this limit is most likely never reached as it requires unreasonable assumptions for the amplification of magnetic fields.

The background magnetic field can be more accurately estimated from a standard model of stellar magnetic fields, that is, a dipolar structure below the Alfv\'en point and a primarily toroidal structure at large distances \cite[e.g.][]{usov1992}.
The most powerful O stars in \ob (e.g. Schulte~7, Schulte~11) have wind terminal velocities around 2500--3000$\kms$, mass-loss rates around $5 \times 10^{-6}\Msyr$, stellar radii around 15--20\,$R_\odot$ and projected rotational velocities around $100\mbox{--}150\kms$ \citep{Berlanas2020}. Assuming a strong surface field ${\sim} 1$\,kG, the limit on the maximum energy is at most 300~TeV. To reach energies above 1\,PeV requires a surface field higher than 10\,kG which is arguably not realistic, in particular since high magnetic fields are correlated with low rotational velocities \citep[e.g.][]{grunhut17}. For Wolf-Rayet stars, the mass-loss rate is typically enhanced by a factor ten but the stellar radius is only a few solar radii \citep{sander2019}. Single Wolf-Rayet stars seem also to be slow rotators \citep{Meynet2003}. Assuming a stellar rotation velocity ${\sim}200\kms$, surface magnetic field ${\sim} 1$\,kG and stellar radius ${\sim} 5\, R_\odot$, the limit on the maximum energy is 200\,TeV. In the case of the very powerful star WR144, which dominates the power output in Cygnus~OB2 with a mechanical luminosity over $9 \times 10^{37}\ergs$, but whose radius is only about $1\,R_\odot$ \citep{sander2019}, the maximum energy falls short of the PeV regime even for extremely large surface fields or rotational velocities. 

On top of the background field, cosmic-ray-induced instabilities are believed to amplify magnetic field fluctuations upstream of strong shocks \cite[e.g.][]{BellLucek, Bell2004}. However, the efficiency of these instabilities is limited in low density environments. For instance, the non-resonant streaming instability at saturation is argued to generate a turbulent field $\delta B_\ur{sat}^2  \approx 4 \pi \xi_\ur{cr}\, \rho V_\ur{w}^3/c$ which, for an acceleration efficiency $\xi_\ur{cr} = 0.1$, a density $\rho = 0.01 m_\ur{p}\pccm$ and a  wind velocity $V_{\rm w} = 3000\kms$, gives $\delta B_\ur{sat} = 4$\,\textmu G which falls short of what would be necessary to accelerate PeV particles.
These arguments dismiss individual stellar winds as well as compact clusters (NGC~6910, NGC~6913) as sources of the ultra-high energy photons observed from Cygnus. We stress that these fundamental arguments apply independently of the boundary conditions downstream of the stellar wind and therefore cannot be circumvented by invoking downstream confinement effects (e.g. via wind-wind collisions, \citealt{bykov2013}). Based on the argument in the previous paragraph, only \ob, with a mechanical power well above $10^{38}\ergs$, could conceivably accelerate PeV particles, though this requires the association as a whole is able to generate a collective large-scale wind termination shock. However, \ob is not a compact cluster: its stellar distribution extends over a diameter of 30~pc with no clear sign of mass segregation. The mechanical power is therefore diluted in a large region and direct wind-wind interactions are not efficient enough to generate a large-scale cluster wind termination shock, as demonstrated by \citet{Vieu2024CygnusSimu}. As such, \ob should be seen as a loose collection of individual stellar winds embedded in a low-density superbubble powered by thermalisation of these stellar winds. The only possible sites of particle acceleration are the individual wind termination shocks around the most powerful O and Wolf-Rayet stars, and these cannot accelerate PeV particles, at least not within our current understanding of microphysical processes near non-relativistic shocks.

In summary, the most likely sources of high-energy ($100\,\ur{MeV} < E < 100$\,GeV) to very high-energy ($100\,\ur{GeV} < E < 100$\,TeV) diffuse $\gamma$-ray emission in the Cygnus region are the stellar winds of the most powerful O and Wolf-Rayet stars, in particular stars lying at the heart of Cygnus~OB2, which can provide a stationary supply of non-thermal particles over millions of years. 
On the other hand, at their current stages of evolution, none of the known sources can reasonably account for the ultra-high energy ($E > 100$\,TeV) emission observed by LHAASO. The ultra-high energy accelerator must therefore be either hidden or transient.

\section{Particle acceleration in a past supernova remnant}
\label{sec:SNRsimulation}

\subsection{The case for a recent supernova in Cygnus~OB2}
\label{sec:sn-evidence}

\begin{figure}
    \centering
    \includegraphics[width=1\linewidth]{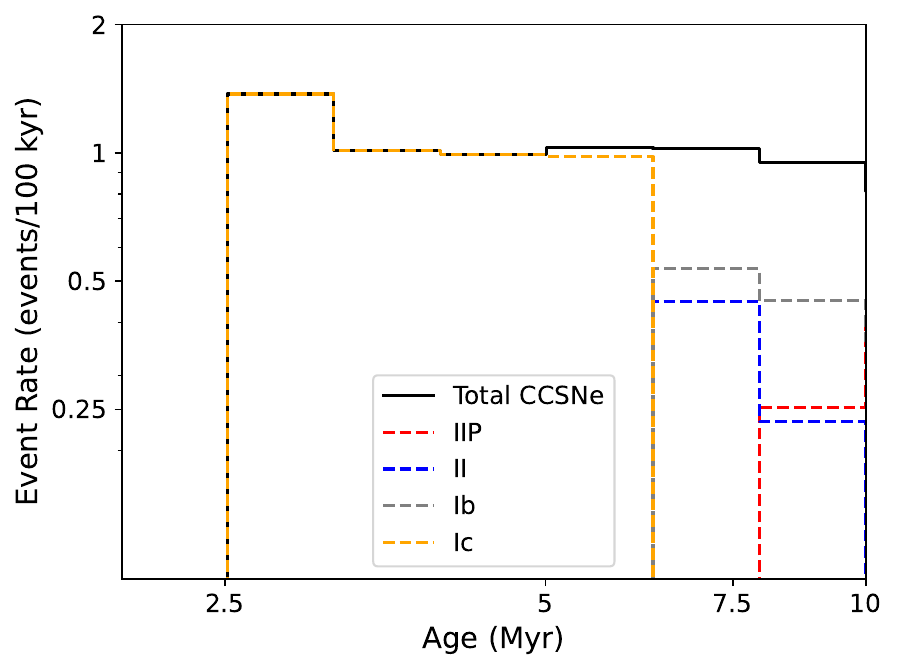}
    \caption{Event rates for different types of core-collapse supernovae for a cluster with a total initial mass of $1.65\times 10^4 \Msun$, as appropriate for \ob, computed using \texttt{Hoki}.}
    \label{fig:SN_Rate}
\end{figure}


Cygnus~OB2 is likely not coeval. Age estimates for its sub-populations range from 3 to 7\,Myr \citep[see e.g.][and references therein]{wright15}. The presence of three Wolf-Rayet stars, 24 supergiants and one pulsar is evidence that late phases of stellar evolution have already been reached and it is likely that supernovae have exploded in the past. To make this argument more quantitative, we show an example calculation of the expected supernova rate using the \texttt{Hoki} package \citep{Stevance2020}, which is an interface for models computed with the Binary Populations And Spectral Synthesis (BPASS) code \citep{Stanway2018}. BPASS models calculate stellar evolution tracks for a wide range of initial parameters, include binary evolution effects, and allow for the calculation of transient rates such as supernovae. For the calculation of supernova event rates, we assume Solar metallicity and a single stellar population with a total initial mass of $1.65\times 10^4 \Msun$ \citep{wright15}. We assume an initial mass function index $\alpha = -2.0$. This is steeper than that suggested for the massive stellar population by \cite{wright15}, but given that the known OB population has increased by $\sim 25\%$ since that estimate \citep{Berlanas2018} this value is reasonable. The calculated rates are shown in Fig. \ref{fig:SN_Rate}. For an age of 3--5\,Myr, type Ic supernovae are by far the most likely type of supernova, and the rate is consistent with several supernovae per Myr. Type Ic supernovae are likely to exceed the canonical supernova power of $10^{51}$\,erg by a factor of a few \citep{Fryer_2018}. The assumption that a powerful supernova has occurred in the last few tens of kyr in Cygnus~OB2 is thus well motivated.

Despite indirect hints of past explosions (e.g. the pulsar PSR J2032+4127 located in \ob, \citealt{Lyne2015}, or runaway stars, \citealt{Comeron2007}) no supernova remnant signature has been confirmed in the close vicinity of \ob, as discussed in Sect.~\ref{sec:candidate_accelerators}. This is perhaps unsurprising. A supernova remnant expanding in the low-density environment carved by the stellar winds around the core of \ob would quickly reach a radius of several tens of parsec and fade below the detection capabilities of radio, optical, and X-ray observatories after a few ten-thousand years \citep{Chu1997}. Besides, the shock is never expected to enter the radiative phase but rather slow down to subsonic velocities after a few 100 thousand years \citep{parizot2004}. 
For these reasons, old supernova remnants expanding in hot bubbles are only expected to leave extended, diffuse, and faint signatures. These would be difficult to detect, in particular in a complex, inhomogeneous environment with, as in the case of Cygnus, a superposition of extended multi-wavelength signals along the line-of-sight.

Assuming a supernova exploded a few tens of thousands of years ago in the \ob association, the maximum energy of particles accelerated at the forward shock is estimated as \citep{vieu2022,vieu2022Emax}:
\begin{align}\label{maxp_SNR}
    E_{\rm max} \approx \left(\frac{B}{5~\text{\textmu G}}\right) \left(\frac{V_\ur{SN}}{10^4 ~\text{$\kms$}}\right) \left( \frac{M_{\rm ej}}{1 M_\odot} \right)^{\frac{1}{3}} 
    \left( \frac{n_\ur{SB}}{0.01 {~{\rm cm}^{-3}}} \right)^{-\frac{1}{3}} \text{~PeV} \, .
\end{align}
This estimate assumes that the remnant expands in a low-density cavity powered by a loose OB association (in contrast to the case of a compact star cluster). For loose associations, the maximum energy reaches at most a few PeV for powerful supernova remnants \citep[see also][]{vieureville2023}. Yet, we show in the following that this can be sufficient to account for the steep component of the ultra-high energy $\gamma$-ray spectrum.

\subsection{Hydrodynamic simulation of a supernova remnant in Cygnus~OB2}
\label{sec:simu}
To estimate quantitatively the maximum energy of particles accelerated by a supernova remnant shock in \ob, we require a prescription for the dynamical evolution of the remnant, accounting for the inhomogeneous superbubble properties ahead of the shock. To this end, we explode a supernova in the previously published hydrodynamic simulation of a \ob-like association \citep{Vieu2024CygnusSimu}. The simulation is resumed at 1.3 Myr, at which point the high-mass supernova progenitor is turned into a Wolf-Rayet star with $\dot{M} = 10^{-5}\Msyr$ and $V_\ur{w} = 2500\kms$. The supernova is then initialised at 1.6\,Myr with a linear radial velocity profile and a flat density profile with a power-law cut-off $\rho \propto r^{-9}$, as described in \citet{Whalen2008} and \citet{Badmaev2024}. The explosion energy is set to $5 \times 10^{51}$\,erg and the core velocity to 16\,000$\kms$, values consistent with a powerful type Ic supernova \citep{Fryer_2018,Barbarin2021}. 
We neglect thermal conduction and cooling, which are not expected to affect the early phases of expansion.

\begin{figure*}
    \centering
    \includegraphics[width=0.85\linewidth]{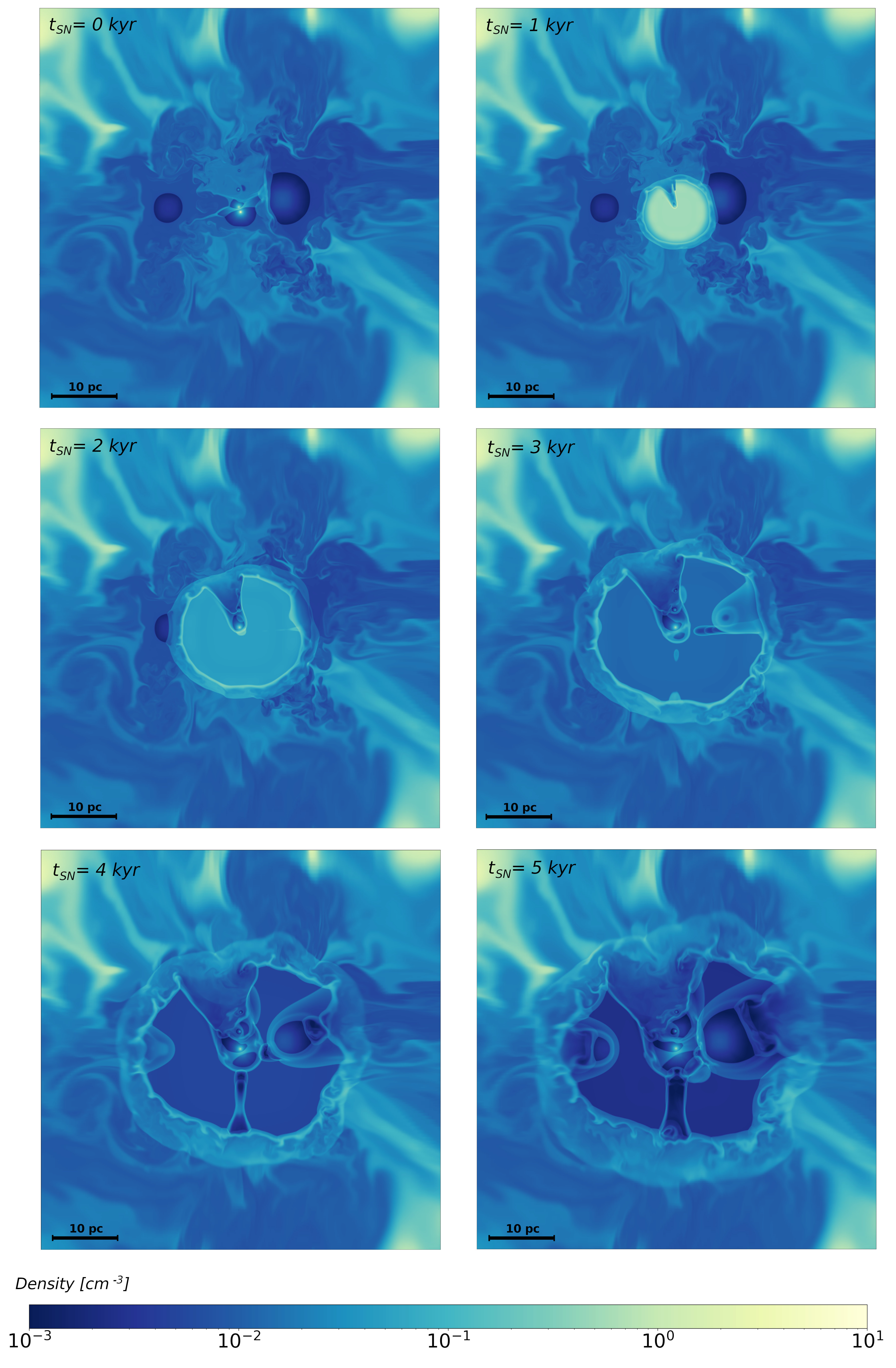}
    \caption{Density slices from the simulation of a supernova remnant expanding in \ob. For details, see Sect.~\ref{sec:simu}.}
    \label{fig:SNR_density_slices}
\end{figure*}

The forward shock of the simulated supernova remnant expands beyond the inner core of \ob, as shown in Fig.~\ref{fig:SNR_density_slices}, and is progressively decelerated. Interaction with the pre-supernova medium, in particular with powerful Wolf-Rayet winds, slows down the expansion further along some directions, producing asymmetries, though the shock expansion is not blocked by the stellar winds. The remnant expands well-beyond the inner core after several thousand years. As the density of the background medium is low (about 0.01~cm$^{-3}$), the free-expansion phase proceeds for a long time, up to about 2~kyr, providing favourable conditions for particle acceleration.

\subsection{Maximum energy of accelerated particles}
\begin{figure*}
    \sidecaption
    \includegraphics[width=12cm]{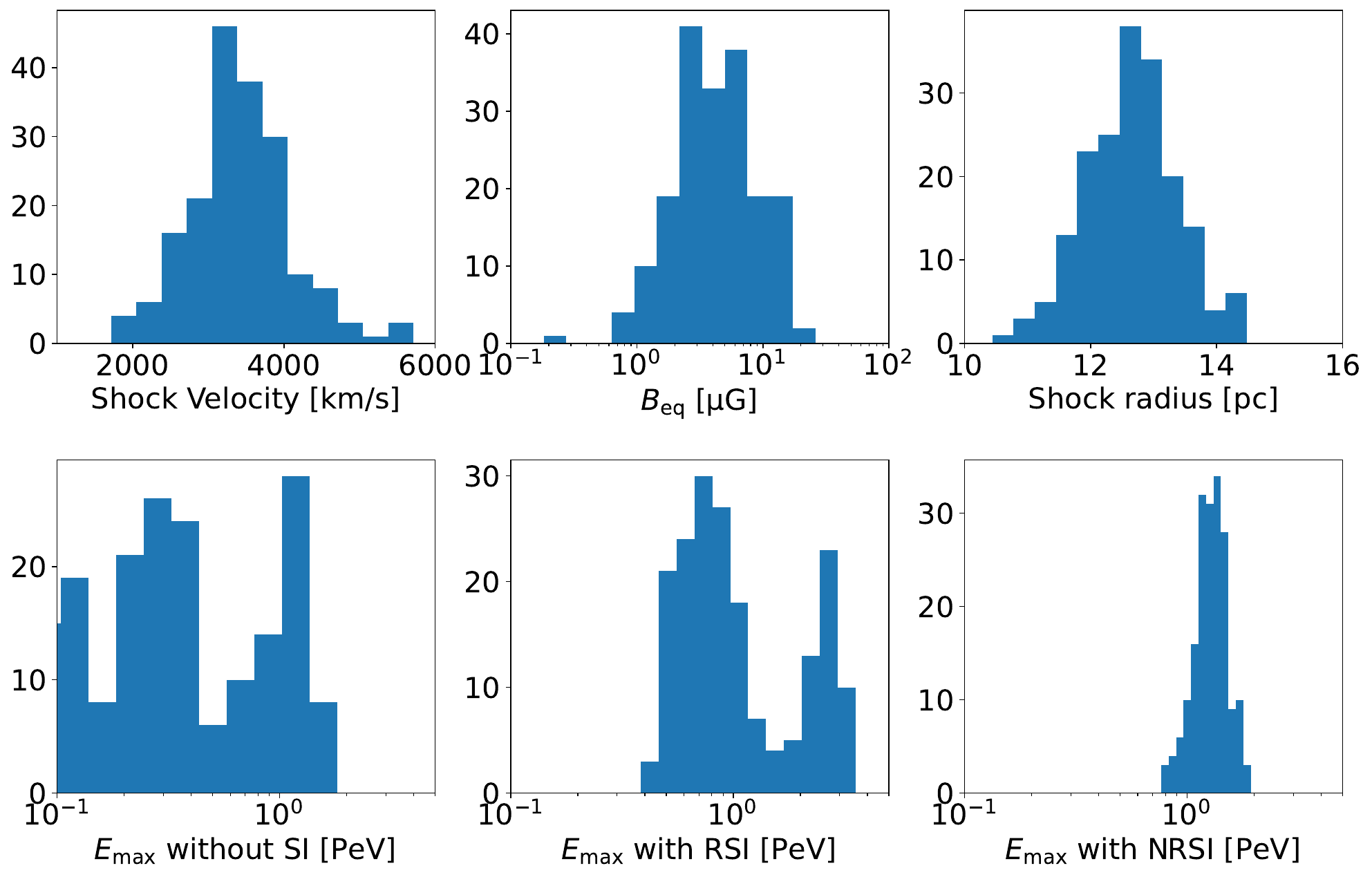}
    \caption{Top: supernova remnant properties across the forward shock surface at 2\,kyr, extracted from the hydrodynamical simulation shown in Fig.~\ref{fig:SNR_density_slices}. Bottom: integrated maximum energy at 2\,kyr, for the resonant and non-resonant streaming instabilities (RSI and NRSI, respectively) and without streaming instabilities. $B_\ur{eq}$ was computed assuming 10\% of the kinetic energy in the superbubble goes into a turbulent magnetic field.}
    \label{fig:SNR_histograms}
\end{figure*}

The simulation was analysed to track the properties of the forward shock as a function of time. The top row of Fig.~\ref{fig:SNR_histograms} illustrates how the properties are distributed across the shock surface at 2\,kyr. At this point, the shock has reached a radius of about 15\,pc and the velocity has decreased to about $3000\kms$, such that diffusive shock acceleration is no longer effective at the required energies. In order to infer the maximum energy of particles accelerated up to this time, we integrate the acceleration rate in time. Since the upstream magnetic field is not calculated in the hydrodynamic simulation, we assume that 10\% of the kinetic energy in the superbubble goes into turbulent magnetic energy, defining $B_\ur{eq} = u\sqrt{4 \pi \eta_\ur{B} \rho}$ with $\eta_\ur{B} = 0.1$ and $u$ the pre-supernova velocity field in the superbubble. Cosmic-ray induced instabilities are further expected to enhance the turbulent component of the background magnetic field and generate a flat turbulence spectrum, although these instabilities grow slowly in a low-density medium. In particular, the non-resonant streaming instability operates only if the cosmic-ray current is strong enough to overcome the magnetic field tension. This condition reads \citep{Bell2004}
\begin{equation}\label{conditionNRSI}
 \xi_\ur{cr} u^3 > v_\ur{A}^2 c\, ,
\end{equation}
where $u$ is the shock velocity, $v_\ur{A}$ the upstream background Alfv\'en speed, and $\xi_\ur{cr}$ is the acceleration efficiency that we set to 10\%. 
The background field, $B_\ur{eq}$, inferred from the simulation is typically about 1--10\,\textmu G. These values are consistent with those we inferred from magnetohydrodynamic simulations in \citet{Haerer2025}. For such $B_\ur{eq}$, the condition \ref{conditionNRSI} is satisfied most of the time upstream of the supernova remnant shock. Provided the streaming instabilities operate, the saturated magnetic field can be estimated as \citep[e.g.][]{BellLucek, Bell2004,matthews2017}
\begin{equation}
    \delta B^2_{\rm sat} \approx \left\lbrace \begin{array}{cc}
      2 \xi_\ur{cr}  u \sqrt{4 \pi \rho} \, B_\ur{eq}    &  \enspace  \mbox{RSI} \\
      4\pi \xi_\ur{cr}\, \rho u^3/c   & \enspace \mbox{NRSI}
    \end{array}
    \right. \, ,
\end{equation}
where RSI denotes the resonant and NRSI the non-resonant streaming instability. In all cases we assume that diffusive shock acceleration operates at maximum efficiency (Bohm diffusion), in which case the acceleration rate is written as \citep{lagage1983}
\begin{equation}\label{accrate}
    \frac{\dd p}{\dd t} \approx \frac{u^2 p}{20 \kappa} \, , \quad \kappa = pc/(3 q B) \, ,
\end{equation}
where $B$ is the turbulent magnetic field. We repeat the calculation assuming first that streaming instabilities do not operate ($B=B_\ur{eq}$), then that RSI operates ($B=\delta B_\ur{sat,RSI}$) and finally that NRSI operates ($B=\delta B_\ur{sat,NRSI}$ if the left hand-side of Eq.~\ref{conditionNRSI} is ten times the right hand-side, $B = B_\ur{eq}$ otherwise).
We integrate the acceleration rate as function of time along a large number of independent lineouts, reading the local shock velocity, density and inferred magnetic field at every timestep. An implicit assumption of this procedure is that particles do not propagate along azimuthal directions from one lineout to another. This is a reasonable assumption if the magnetic field is parallel to the shock normal. At every time-step we check that the Hillas confinement limit, $E < q B R u/c$, where $R$ is the distance travelled by the forward shock, is not violated, otherwise integration is not performed. Efficient particle acceleration stops after about 2\,kyr when the supernova remnant enters the Sedov-Taylor phase and the highest energy particles are no longer effectively confined.

The second row of Fig.~\ref{fig:SNR_histograms} shows the maximum energy obtained for protons at the end of the phase of efficient particle acceleration. Without streaming instabilities, the supernova remnant shock is overall able to accelerate efficiently up to about 500\,TeV. With streaming instabilities, a maximum proton energy of a few PeV is reached. Note that we naively expect the maximum energy to scale as the square root of the explosion energy such that a nominal supernova ($10^{51}$\,erg) is expected to accelerate protons up to 0.5--1\,PeV with streaming instabilities.

In the approach adopted above, we chose optimal parameters for confinement and acceleration of the highest energy particles, as required to reach the Hillas limit. 
This is usually motivated by the fact that cosmic rays self-excite the turbulence. Independently of this mechanism,
studies by \citet{bell25} \citep[see also][]{2021ApJ...923...53L} find that particle mirroring in large-scale fields -- those larger than the gyroscale of the accelerating particles -- can exceed quasi-linear theory estimates of particle confinement near shocks, even with moderate (sub-Bohm) scattering rates. 
Large-scale field structures naturally result from the turbulent interactions of magnetized stellar winds within superbubbles \cite[see for example][]{Haerer2025}. These findings provide confidence that the Hillas limit could be achieved without resorting to the Bohm scattering limit.

\section{Molecular Gas}
\label{sec:gas}
\label{sec:clouds}

In this section, we reconstruct the 3D molecular gas distribution in the Cygnus region from previous studies. This is subsequently used to compute the $\gamma$-ray emission (see Sect.~\ref{sec:modelling} and Sect.~\ref{sec:gamma-sn}).

CO intensity maps reveal a huge molecular complex from about 76$^\circ$ to about 85$^\circ$ in longitude and from about -3$^\circ$ to about +4$^\circ$ in latitude. Within this complex, five regions are especially prominent: the cloud in the far-north which separates the North-America and the Pelican nebulae, the northern complex between 80--83$^\circ$ (Cygnus-X North), the southern complex between 77--80$^\circ$ (Cygnus-X South), the north-west clouds at high latitudes and the lane extending in the far-south, including the S106 region. It has been established that this molecular complex is not a coherent structure but a superposition of clouds along the line-of-sight, including the well-known Cygnus Rift in the foreground \citep[e.g.][]{Reipurth2008}.

\citet{Schneider2006} established the first comprehensive census of the molecular clouds in Cygnus-X, using $^{12}$CO and $^{13}$CO emission lines. Because the Cygnus region is seen tangentially to the galactic rotation, kinematic velocities do not translate into reliable distance measurements. Restricting velocities to the range $(-20,30)\kms$ allows to exclude the Perseus Arm and beyond, however, the 3D distribution of the gas in the Local Arm (within 3\,kpc) cannot be directly modelled. Nevertheless, scanning the intensity map along the velocity axis 
can give some information about superimposed components, at least at the qualitative level. The underlying assumption is that coherent structures correlate in position-position-velocity space.

\begin{figure*}
    \centering
    \includegraphics[width=\linewidth]{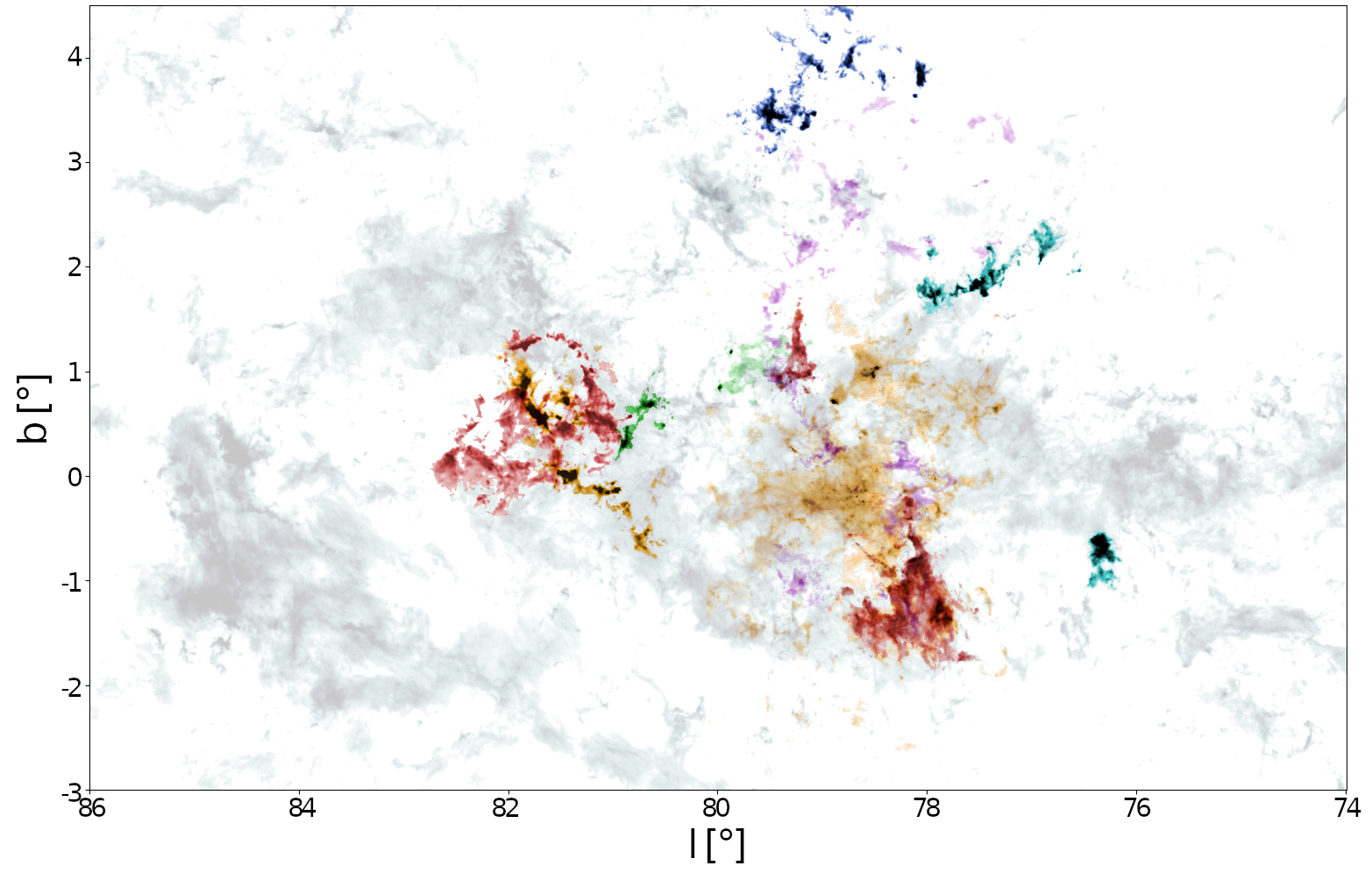}
    \caption{\label{fig:CygnusClouds}Main molecular clouds in the Cygnus region.
    Red: Cygnus 1.3 North at $l>80^\circ$ (Schneider Group II, 1.3\,kpc) and Cygnus 1.3 South at $l<80^\circ$ (part of Schneider Group IV, $1.3$\,kpc). Orange: Cygnus North Filament at $l>80^\circ$ (Schneider Group I, 1.5\,kpc) and Cygnus SFR South at $l<80^\circ$ (part of Schneider Group IV, unknown distance ${>}1.4$~kpc).
    Pink: Cygnus South Filament (part of Schneider Group IV, unknown distance ${>}1.4$~kpc).
    Green: Cygnus Central Clumps (Schneider Group III, $1.6\mbox{--}1.8$\,kpc). Dark blue at high latitude: Cygnus West, unknown distance (${>}1.4$\,kpc). Turquoise: S106 regions (described in \citealt{Schneider2007}, $1.6\mbox{--}1.8$\,kpc). The grey background shows foreground clouds: the North-America and Pelican nebulae on the left, the North-West clouds, and the Rift lane extending to the right in the direction to the Galactic Centre.}
\end{figure*}

\begin{table*}
    \centering
    \caption{Molecular clouds in the Cygnus region.}
    \begin{tabular}{llllllll}
         & Name                  & l [$^\circ$]   & Distance [kpc]  & Distance ref         \\
        \hline
        A & Foreground Rift       & 74--86   & 0.7--1.1                     & \citet{Zhang2024}   \\
        B & Cygnus North-West     & 79.5--81 & ${\sim}1$                     & \citet{Zhang2024}     \\
        C & Cygnus 1.3 North      & 80--83.5 & 1.3--1.4                      & \citet{Zhang2024}     \\
        D & Cygnus 1.3 South      & 77--80   & 1.2--1.3                      & \citet{Zhang2024}     \\
        E & Cygnus North Filament & 80.5--82 & 1.4--1.5                       & \citet{Rygl2012}      \\
        F & Cygnus Central Clumps & 79--81   & 1.6--1.8                       & \citet{Orellana2021,Xu2013}        \\
        G & Cygnus West           & 78--80   & ${>}1.4$                         &                      \\
        H & Cygnus SFR South      & 77--80   & ${>}1.4$                       &                      \\
        J & Cygnus South Filament      & 77--79   & ${>}1.4$                        & \\
        K & Cygnus Far South      & 76--78   & 1.6--1.8                     & \citet{Schneider2007} \\
        \hline
    \end{tabular}
    \tablefoot{Section~\ref{sec:clouds} describes the procedure by which the clouds were identified. Figure~\ref{fig:CygnusClouds} shows a map of the clouds. The specific distance and mass estimates assumed in the calculation of the $\gamma$-ray flux are provided in \url{https://doi.org/10.17617/3.G9LUS1}.}
    \label{tab:MolecularClouds}
\end{table*}

After carefully scanning all prominent structures in the CO map, \citet{Schneider2006} extracted four seemingly coherent groups, in addition to the foreground Rift. Group I is a filamentary cloud north of Cygnus which hosts several dense clumps forming protoclusters seen as radio sources, for example DR21, DR20, DR22, and DR23 around $l=81^\circ$ in the velocity range $(-7,1)\kms$. In the following we refer to this group as the ``North Filament''. Group II partly overlaps with the filament but is significantly decorrelated in velocity space with a range $(6,20)\kms$, and was placed in the foreground by early absorption studies \citep{Dickel1978}. Parallax measurements of masers \citep{Rygl2012} place DR20 and DR21 (North Filament) at about 1.5\,kpc, and group II, which includes in particular W75N and DR17, in the foreground at about 1.3\,kpc \citep[see also][]{Gottschalk2012}. 
Group III gathers compact clumps and globules around Cygnus~OB2, notably DR18 and the southern arm of DR20. The star-forming region AFGL 2591 was originally included in this group, however it has since been relegated to the background (about 3.3 kpc) by a maser measurement \citep{Rygl2012}. Finally, most of the southern clouds were gathered in a single Group IV, and the remaining unassociated features were attributed to the foreground Cygnus Rift (Group V) between $0.6\mbox{--}0.8$\,kpc. \citet{Schneider2007} completed the census in the southernmost part up to the S106 HII region and showed that several substructures in this area are shaped by stellar feedback in a star-forming region at $1.6\mbox{--}1.7$\,kpc, which hosts notably the compact star clusters NGC~6913, Berkeley~86 and Berkeley~87.

\medbreak

The groups identified by \citet{Schneider2006} can serve as a basis to identify substructures in Cygnus, however the relative masses of the groups are highly uncertain and accurate distances cannot be directly inferred. In addition, the assumption that coherent structures in three-dimensional space correlate in position-position-velocity space is doubtful. Turbulence can dramatically spread the velocity component across a given molecular cloud, especially near active star-forming-regions, which leads to over-decomposition of coherent clouds. On the other hand, overlapping clouds can have similar velocities by coincidence, which leads to confusion between the various components along the line-of-sight. The advent of the {\textit{Gaia}} era, enabling precise parallax measurements, has shed new light onto the structure of the molecular clouds in Cygnus. \citet{Zhang2024} employed a clustering algorithm to identify CO clouds and measured their distance via extinction jumps along the line-of-sight.
This successfully disentangles foreground components, showing that the Rift actually extends up to 1.1\,kpc and is much more massive than thought by \citet{Schneider2006}. In particular, it becomes clear that the Group IV in \citet{Schneider2006} is largely confused with the foreground Rift. In addition, \citet{Zhang2024} measure a massive cloud in the south with a distance estimate of 1.3~kpc, including DR13 and its prominent filament (red triangular cloud in Fig.~\ref{fig:CygnusClouds}). This brings a southern counterpart to Schneider Group II. We propose to name these two groups in the 1.3\,kpc layer ``Cygnus 1.3 North'' and ``Cygnus 1.3 South'' for clarity.

\subsection*{Cygnus-X North}

Unfortunately, the extinction becomes too high beyond ${\sim}1.4$\,kpc, preventing \citet{Zhang2024} to probe more distant molecular clouds, and in particular the clouds in the vicinity of the star-forming region at ${\sim}1.6\mbox{--}1.8$\,kpc (including Cygnus~OB2).
In Cygnus North, \citet{Zhang2024} were able to associate a distance to about 50\% of the total mass, which represents about 115 thousand solar masses in the foreground of the North Filament. On the other hand, \citet{Schneider2006} associated a mass of 167 thousand and 55 thousand, respectively for the North Filament (Group I) and Cygnus 1.3 North (Group II) which, corrected for the updated distance measurements, adds up to about 160 thousand solar masses. Within mass inference uncertainties, it is therefore reasonable to assume that the census is complete in the north, with Cygnus 1.3 North and the North Filament lying behind the Rift. Using the $^{12}$CO data from \citet{Zhang2024}, we extract the shape of the North Filament and Cygnus 1.3 North in the velocity ranges $(-7,1)\kms$ and $(7,15)\kms$, respectively. These are shown in red and orange in Fig.~\ref{fig:CygnusClouds}. Note that the most recent distance estimates place Cygnus~OB2 at about 1.65~kpc, which is behind the North Filament (1.5\,kpc). The North Filament might actually lie at the edge of the putative \ob superbubble, which could explain its peculiar shape \citep{Li2023}.

\subsection*{Cygnus-X South}

The situation is less clear in the south. \citet{Zhang2024} were able to associate a distance to approximately 150 thousand solar masses, corresponding to only 36\% of the total mass. The missing mass most likely lies beyond 1.4\,kpc, possibly embedded in the interfaces of photodissociation regions at the distance where several active star-forming regions are located, beyond the reach of the extinction jump method (${>}1.3$~kpc). We suggest to call this hidden group ``Cygnus SFR South''. In order to extract this group, we identify regions of the position-position-velocity space which do not correspond to the clouds catalogued by \citet{Zhang2024}.
These regions appear in the velocity ranges $(-10,-5)\kms$, $(2.2,4.5)\kms$ and $(4,8.5)\kms$ around $78\mbox{--}79 ^\circ$ in longitude. 
The result defines the ``Cygnus SFR South'' region shown in orange at $l<80^\circ$ and $-1^\circ \lesssim b \lesssim 1^\circ$ in Fig.~\ref{fig:CygnusClouds}. In the range $(12,20)\kms$, a filamentary cloud appears, which we call the ``South Filament'' (shown in pink in Figure~\ref{fig:CygnusClouds}). Finally, another region at high latitude, $b \approx 4^\circ$ in the velocity range $(-5,3)\kms$ is extracted and labelled ``Cygnus West''. We assume in the following that all these southern clouds beyond the reach of the extinction jump method are located within the active star-forming region at about 1.7\,kpc, loosely coincident with the photodissociation regions. This aligns with a plausible scenario where these clouds are at the intersections of multiple cavities that were carved in the past by the joint feedback of \ob, NGC~6913, NGC~6910, and possibly other clusters and associations (see Sect.~\ref{sec:intro}).

\subsection*{Central clumps}

In addition to the massive molecular clouds in the north and south, a number of clumps, sometimes called ``globules'' or ``condensations'', are seen around \ob, with hints of direct interaction with the radiative feedback from the association \citep[e.g.][]{schneider16}. A number of these clumps are listed in Table 3 of \citet{Zhang2024}, to which we further add the prominent regions DR18 and DR20 in the velocity range $(-11,1)\kms$ around $l=81^\circ$. We group these clumps under the common label ``Cygnus Central Clumps'' and assume a distance similar to that of Cygnus~OB2 (1.6\,--1.8\,kpc).

\subsection*{Other groups}
The remaining foreground groups (North-West, 1.3 South) can be directly identified from the distance measurements given in \citet{Zhang2024}. Finally, Cygnus Far-South can be extracted manually according to its well-defined shape described in \citet{Schneider2007}. Table~\ref{tab:MolecularClouds} summarises our census of the molecular cloud complexes in the Cygnus region. The listed masses are inferred by assuming a $X_\ur{CO}$ factor of $4 \times 10^{20}$ H/cm$^2$ (K\,km\,s$^{-1}$)$^{-1}$. For later convenience, we introduce shorthands A--K to refer to the clouds.

\section{Modelling particle propagation and $\gamma$-ray emission}
\label{sec:modelling}

In the previous sections, we motivated that a powerful past supernova in Cygnus is a plausible source of very-high energy particles. We estimated the maximum energy of particles accelerated at the forward shock and discussed the 3D distribution of target material in molecular clouds. Here, we outline our approach to model particle transport and $\gamma$-ray emission. We construct both hadronic and leptonic models and apply them to supernova and stellar-wind scenarios in Sect.~\ref{sec:gamma-sn} and \ref{sec:gamma-ob2}.

\subsection{Losses, photon fields, and magnetic field}
\label{sec:phfields}

\begin{figure}
    \centering
    \includegraphics[width=\linewidth]{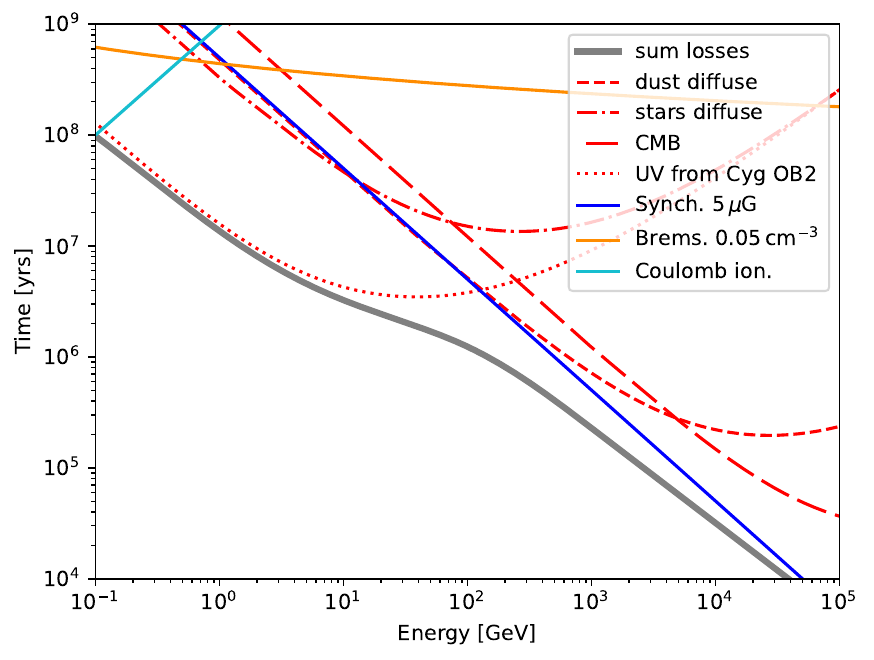}
    \caption{Loss timescales for electrons in the putative superbubble surrounding \ob. The photon field from stars in \ob is evaluated at 20\,pc. Diffuse dust and starlight are taken from the model by \citet{popescu17} and enhanced by a factor of two. Coulomb losses are shown for a hot, low-density, ionised medium ($T=10^6\,$K, $n_\ur{SB}=0.05\pccm$) following \citealt{Schlickeiser02} (Sect.~4.5). All other timescales are computed with the \texttt{GAMERA} code \citep{Hahn2022}.}
    \label{fig:ic-losses}
\end{figure}


Knowledge of ambient photon fields and the magnetic field is required to compute leptonic losses and inverse Compton emission. We model galactic diffuse starlight and dust-scattered starlight following the axisymmetric model of \citet{popescu17}, with an enhancement factor of two for both photon fields, to reflect increased activity within the spiral arm \citep[c.f.][]{Breuhaus2021}. The cosmic microwave background is included as a blackbody with a temperature of 2.7\,K and energy density of $0.26\eVccm$. We model the photon fields of individual massive stars in \ob with blackbody spectra, taking luminosities and temperatures from \citet{wright15}. The total photon field is obtained by summing all spectra. The energy density at a distance $R$ is obtained from the luminosity, $L$, as $L/(4\pi R^2 c)$. At a distance of 20\,pc, the energy density of the cluster photon field is $25\eVccm$, which is comparable to the model value obtained by \cite{schneider16}, who take into account the spatial distribution of stars. Figure~\ref{fig:ic-losses} shows cooling time-scales for electrons, for a magnetic field of 5\,\textmu G and a density of $0.05\pccm$, which is representative of the superbubble interior. We assume that the magnetic field is constant over the superbubble interior. In the GeV range, the dominating photon field is the cluster photon field. At higher energies, losses are dominated by synchrotron cooling. Losses for nuclei are neglected as their timescales are large in the low-density superbubble environment.

\subsection{Target material}
\label{sec:target-gas}

\begin{figure*}
    \centering
    \includegraphics[width=\linewidth]{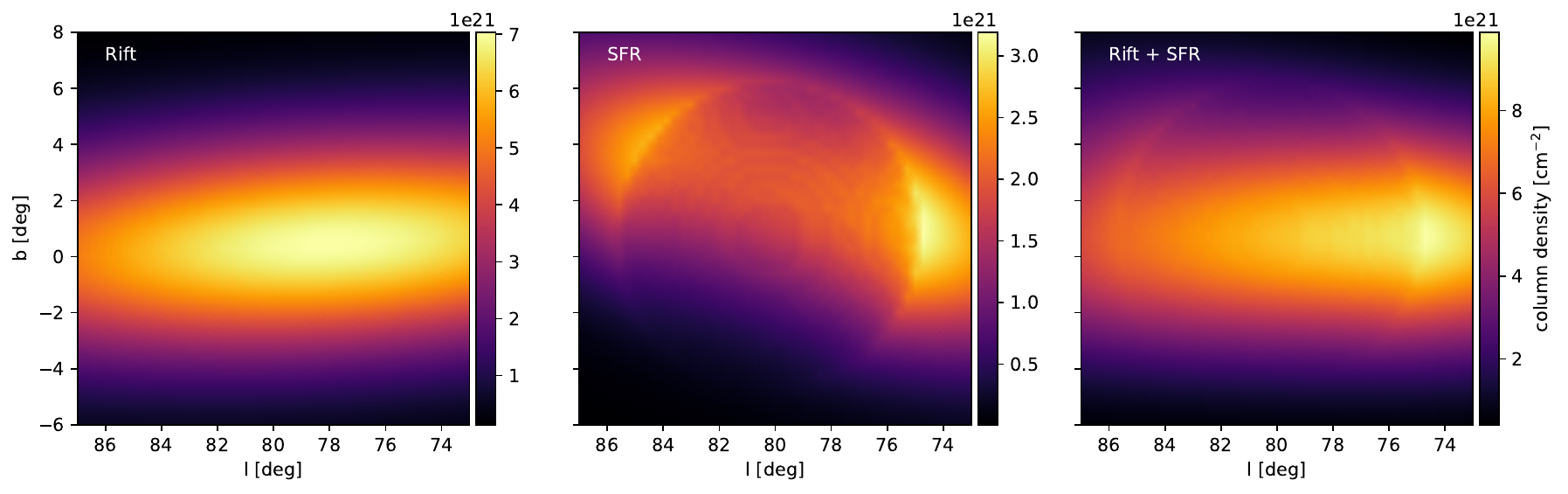}
    \caption{Neutral gas model. The model consists of two 3D-Gaussian components, the Cygnus Rift (left) and the Cygnus SFR (middle), with a spherical (radius 150\,pc), low-density ($0.05\pccm$) cavity around \ob. The total mass is set to match the mass obtained from integrating the neutral hydrogen column density obtained by \citet{Astiasarain2023}. For model parameters, see Tab.~\ref{tab:HI}.}
    \label{fig:HImap}
\end{figure*}

\begin{table}
    \centering
    \caption{Parameters of the neutral gas model}
    \begin{tabular}{c|c|c}
        Parameter  & Rift & SFR \\
        \hline
        centre $d$ [kpc]           & 1.0  & 1.7    \\
        centre $l$  [$^\circ$]   & 78   & 79     \\
        centre $b$  [$^\circ$]   & 0.5   & 2       \\
        total mass [$10^6\Msun$]    & 4.2     & 3.4         \\
        $\sigma_l$ [pc] & 200 & 200 \\
        $\sigma_b$ [pc] & 50 & 80 \\
        $\sigma_d$ [pc] & 200 & 200 \\
        rotation [radian] & 0.05 & 6 \\
    \end{tabular}
    \tablefoot{The components are modelled as 3D Gaussians with standard deviations $\sigma_l$, $\sigma_b$, and $\sigma_d$ in galactic longitude, latitude, and distance. The rotation is measured with respect to the galactic plane. Figure~\ref{fig:HImap} shows column density maps of the model.}
    \label{tab:HI}
\end{table}


We consider proton-proton interactions in molecular clouds and neutral gas. The distances of clouds A--B and C--D in Tab.~\ref{tab:MolecularClouds} are fixed to 1\,pc and 1.3\,kpc respectively. We note that, assuming a distance of 1.6\,kpc for \ob, the density of particles at clouds A--D is too low to materially contribute to the flux. These clouds however contribute to diffuse emission from galactic cosmic rays. Beyond 1.4\,kpc, extinction introduces considerable uncertainty in cloud positions (see Sect.~\ref{sec:clouds}). Clouds E--K are initially assumed to be equidistant to \ob and close to the superbubble edge. This is a plausible scenario: density increases in the superbubble shell, which is likely to be thin and fragmented due to cooling \citep[see][]{weaver1977}. Such a scenario is also plausible considering the CO data (see Sect.~\ref{sec:clouds}). 

We construct a neutral gas model based on measurements of the column density of atomic hydrogen (HI) by \citealt{Astiasarain2023} (see their Fig.~2). We associate the neutral gas with known molecular cloud complexes: the Cygnus Rift at a distance of about 1\,kpc and the Cygnus SFR at about 1.6--1.8\,kpc. Each component is modelled as an ellipsoid with parameters listed in Tab.~\ref{tab:HI}. Model parameters and total masses are chosen to account for the total column density, thus the model assumes that all mass is contained in the two components mentioned above. 

As discussed in Sect.~\ref{sec:intro}, there are a number of observations hinting at the existence of a low-density ionised cavity around Cygnus~OB2, which shapes the nearby molecular clouds \citep[e.g.][]{Li2023}. The shape and full extent of this cavity cannot be directly inferred from observations. According to \citet{weaver1977}, assuming a continuous injection of energy and a homogeneous ambient medium, the radius of such a wind blown bubble is expected to be
\begin{equation}    
    \label{eq:rfs}
    R_\text{b} = 183 \left( \frac{\xi_\mathrm{b}L_\text{w}}{10^{38}\ergs} \right)^{1/5} \left( \frac{n_\ur{ISM}}{1\,\mathrm{cm}^{-3}} \right)^{-1/5} \left( \frac{t}{5\,\mathrm{Myr}} \right)^{3/5} \mathrm{pc} \, ,
\end{equation}
where $L_\ur{w}$ is the wind power, $n_\ur{ISM}$ the ambient density, $t$ the age of the system, and $\xi_\mathrm{b}$ an empirical correction factor \citep[see][]{vieu2022}. Modelling a spherical superbubble is a necessary simplification given the difficulty to observationally determine its real boundaries. Based on the current power, past stellar activity (see Sect.~\ref{sec:ob2}) and observational hints, we set the superbubble radius to 150\,pc from the centre of Cygnus~OB2 ($l=80.2^\circ$, $b=0.8^\circ$). 
Inside the superbubble, we set the density to a constant value of $0.05\pccm$. This value lies in the range typical for superbubbles \citep[][see also Sect.~\ref{sec:intro}]{weaver1977}. We note that the density obtained in the hydrodynamical simulation (see Sect.~\ref{sec:SNRsimulation}) is lower, because thermal conduction is not included and evaporation at the shell is hence omitted.

\subsection{Particle injection and propagation}
\label{sec:prop}

We assume that at the end of the efficient particle acceleration phase, particles are released into the low-density superbubble carved by the winds from massive stars in \ob. In the resulting turbulent environment, the diffusion coefficient, $D_{\rm SB}$, is expected to be smaller than its average interstellar value. We assume $D_{\rm SB} = 10^{25} p_{\rm GeV}^{1/2}$\,cm$^2$\,s$^{-1}$ which, in the framework of quasi-linear theory, is expected for a magnetic field around 10\,\textmu G and a coherence length of a couple of pc. Beyond the low-density superbubble, the diffusion coefficient is increased to its fiducial value in the interstellar medium, $D_{\rm ISM} = 10^{28} p_{\rm GeV}^{0.4}$\,cm$^2$\,s$^{-1}$. 

We model large-scale advection in a mean $r^{-2}$ wind profile with a speed of $100\kms$ at 10\,pc. We note that advection is only relevant at low energies, especially for our chosen initial velocity, which is appropriate for loose associations such as \ob \citep[see][]{Vieu2024CygnusSimu}. For electrons, we take inverse Compton, synchrotron, and bremsstrahlung losses into account, as described in Sect.~\ref{sec:phfields}.
The resulting inhomogeneous transport equation for the particle distribution function $f(r,p,t)$ reads,
\begin{align}
\partial_t f
  & + u \partial_r f
    - \frac{1}{r^2} \partial_r \left(r^2 D \partial_r f\right)
    - \frac{1}{p^2} \partial_p \left( p^2 \frac{\dd p}{\dd t}  f \right)
= Q(r,p,t) \,,
\end{align}
where we adopt spherical symmetry.
Adiabatic losses vanish for our adopted $r^{-2}$ radial velocity profile. $Q(r,p,t)$ is a source term which can represent, for instance, impulsive injection from a supernova remnant or stationary injection from wind termination shocks.
The transport equation is an example of a advection-diffusion-reaction equation. We solve it with a finite element discontinuous Galerkin method, adapting the method presented in \citet[Sect.~4.6]{DiPietro2012Mathematical} for time-dependent problems, using a Crank-Nicolson integrator and an upwind-flux for the advection term.
For efficient computations, we use logarithmic $p$ and $r$ variables. 
We use zero Dirichlet boundary conditions everywhere, suppressing the diffusion term below 10\,pc to avoid propagation to the centre. We made sure that the solution was not impacted by the choice of boundary conditions, especially checking that the outer $r$ boundary was far enough away for the solution to sufficiently decay.

The solver was implemented in C++ using the open-source finite-element library deal.II \citep{2024:africa.arndt.ea:deal,dealii2019design}.
The code was validated against a number of tests for which analytic solutions are known, including homogeneous synchrotron losses in a box, advection-diffusion without losses, and an impulsive Gaussian profile with homogeneous diffusion coefficient. The code provides the solution for $f(r,p,t)$ which can be post-analysed to determine the non-thermal emission.

\subsection{Gamma-ray emission} 
We compute non-thermal emission using the \texttt{GAMERA} code \citep{Hahn2022}\footnote{\url{http://libgamera.github.io/GAMERA/docs/main_page.html}}. In Sect.~\ref{sec:gas}, we provided the details of our model for the 3D distribution of target material, which includes both the molecular clouds and neutral medium. Convolving this with $f(r,p,t)$ yields both the full spectra integrated over the region and synthetic $\gamma$-ray emission maps. For the leptonic model, we evaluate the UV photon field from \ob as a function of radius and keep all other photon fields and the magnetic field strength constant.

We fitted the $\gamma$-ray morphology by adjusting the positions of clouds E--J. Parameters of the neutral gas model were kept fixed, including the superbubble interior density of $0.05\pccm$. Emission from proton-proton interactions in molecular clouds was evaluated at the barycentre of each individual cloud. We evaluated the neutral gas distribution on a 3D grid with cell size 16.8' by 16.8' in $l$ and $b$ and 24\,pc in distance.

The diffuse galactic cosmic-ray background is modelled following \citet{Schwefer2023}. We evaluate their model for atomic and molecular hydrogen and increase the resulting flux by a factor of 1.7 to account for contributions from heavier nuclei \citep{kafexhiu2014}. We assume a distance of 1.6\,kpc to \ob to convert luminosity to flux.

\section{Hadronic emission from a supernova remnant}
\label{sec:gamma-sn}

\subsection{Setup and results}

\begin{figure}
    \includegraphics[width=\linewidth]{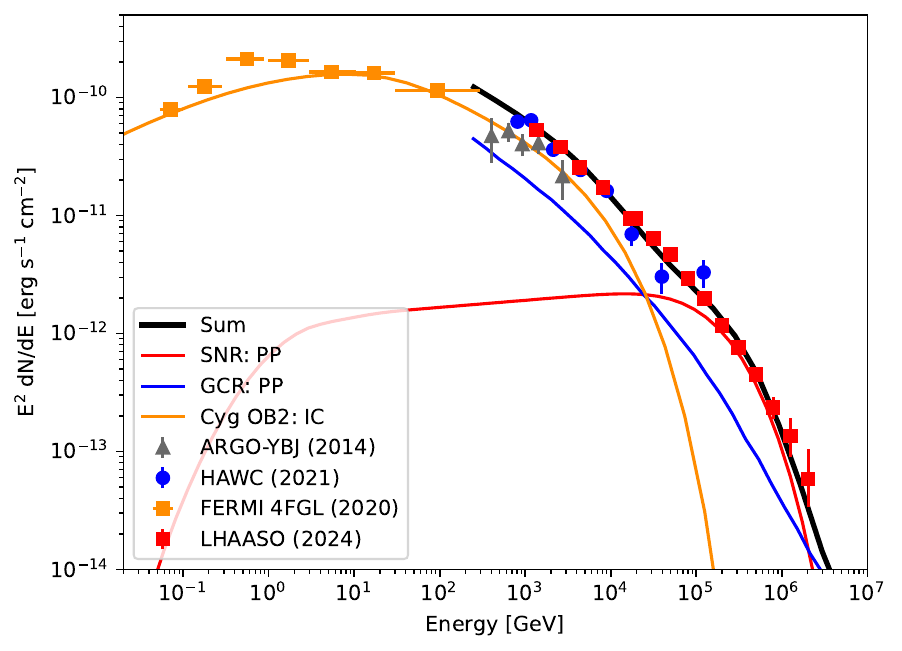}\\
    \includegraphics[width=\linewidth]{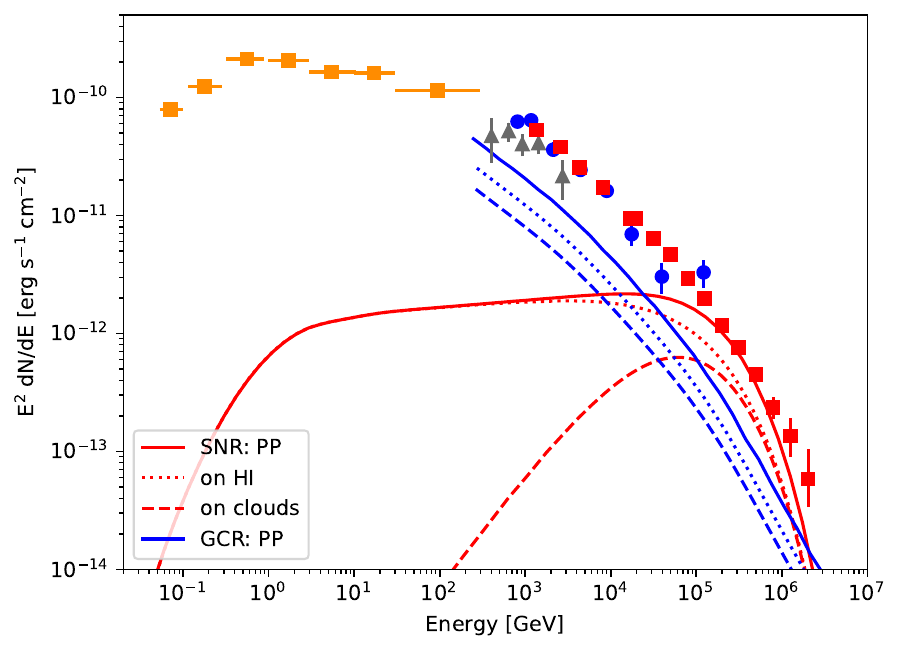}\\
    \includegraphics[width=\linewidth]{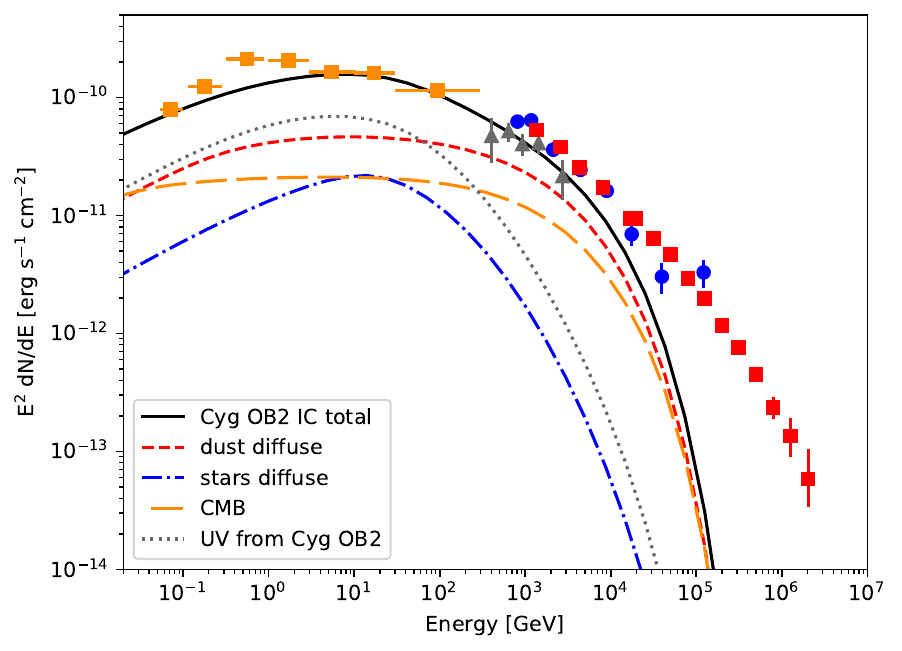}
    \caption{Preferred model for the region-integrated $\gamma$-ray spectrum, consisting of hadronic emission from a 50\,kyr-old supernova remnant along with inverse Compton emission from electrons accelerated at stellar-wind termination shocks in \ob. The top panel shows both components in red and yellow, as well as emission from diffuse galactic cosmic rays in blue. Diffuse galactic cosmic rays are only shown above 300\,GeV, since they are included in the \fmlat background model. The middle panel shows hadronic model components. The lower panel shows contributions from inverse Compton scattering on individual photon fields. Data are taken from \citet{dataArgo2014}, \citet{gammaCygnus_HAWC2021}, \citet{dataFermi4FGL2020}, and \citet{Lhaaso2024}.}
    \label{fig:sed_gamma}
\end{figure}

\begin{figure*}
    \centering
    \includegraphics[height=0.38\linewidth]{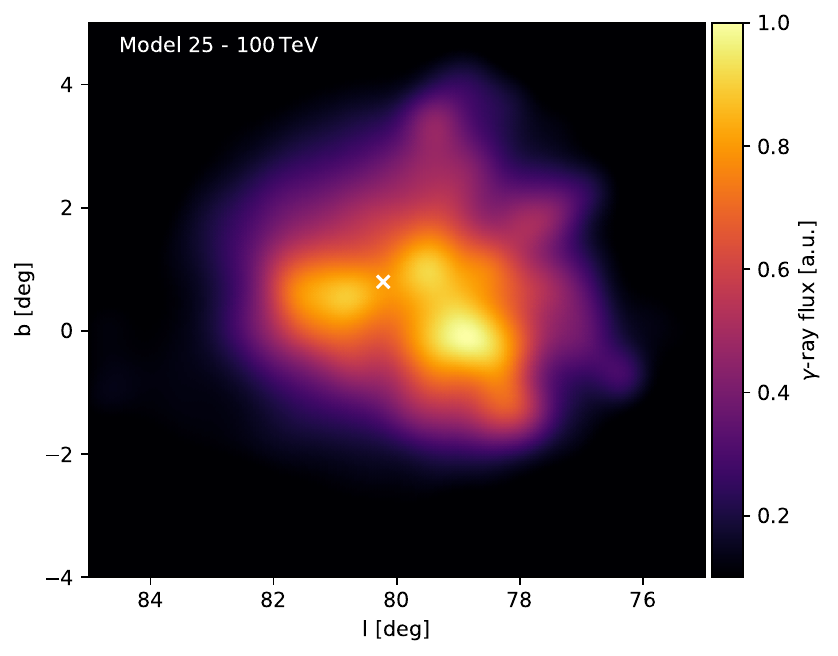}
    \includegraphics[height=0.38\linewidth]{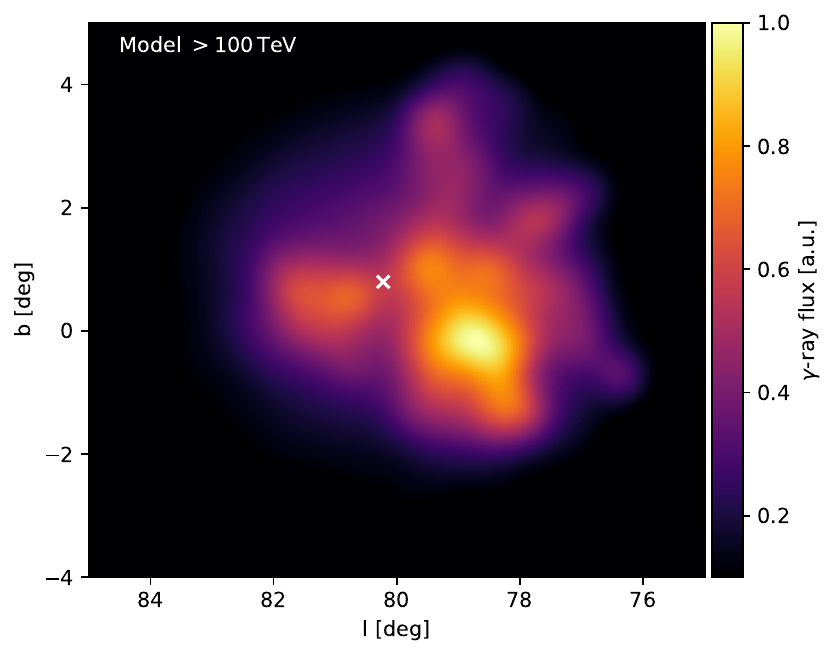}
    \includegraphics[height=0.38\linewidth]{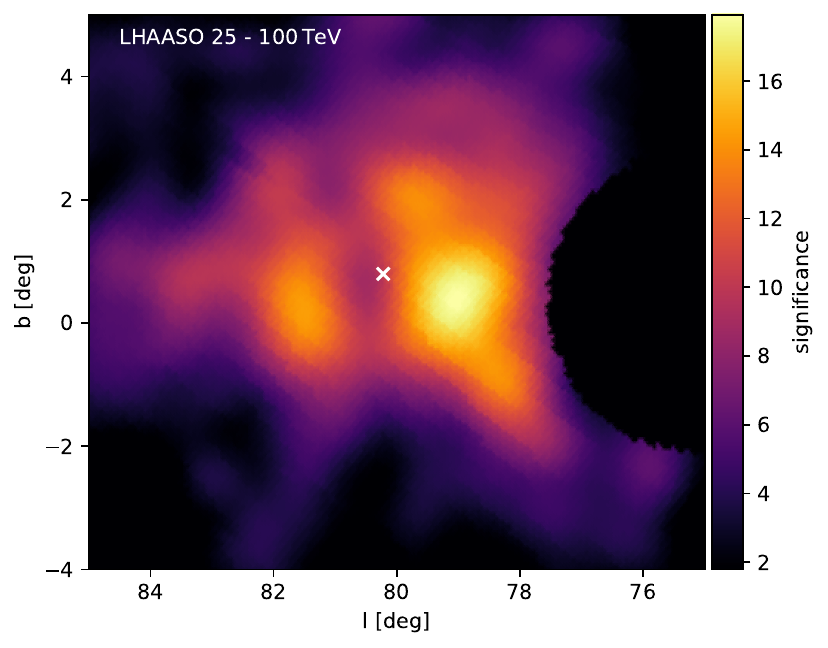}
    \includegraphics[height=0.38\linewidth]{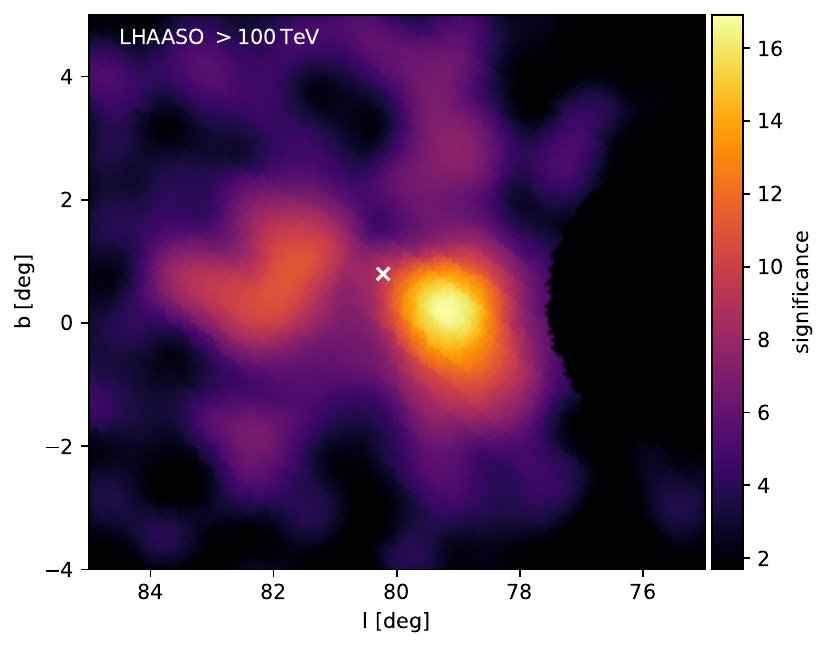}
    \caption{Synthetic $\gamma$-ray emission maps (top row) and significance maps from \citet{Lhaaso2024} (bottom row). Both model and data are smoothed using a Gaussian kernel of $\sigma=0.3^\circ$. The top of the colour scale is set to the maximum and the bottom to 10\% of the maximum. The white cross marks the position of the stellar association \ob. The figure demonstrates that the model is in good agreement with the $\gamma$-ray morphology and its energy dependence.}
    \label{fig:map_gamma}
\end{figure*}

\begin{figure}
    \centering
    \includegraphics[width=\linewidth]{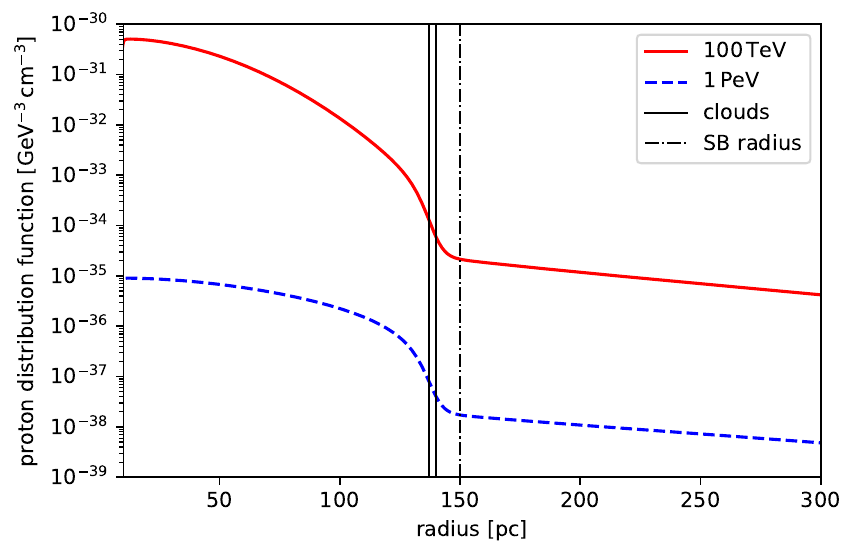}
    \caption{Proton distribution function as a function of radius at 100\,TeV (blue-dashed line) and 1\,PeV (red-solid line). The dash-dotted vertical line marks the superbubble radius, which separates the low-diffusion and interstellar-medium diffusion zones. The solid vertical lines indicate the positions of molecular clouds in the model. The relative normalisation between the two displayed proton distributions changes strongly at small radii, which results in energy dependent morphology (see Fig.~\ref{fig:map_gamma}).}
    \label{fig:fSN}
\end{figure}

\begin{figure*}
    \centering
    \includegraphics[width=\linewidth]{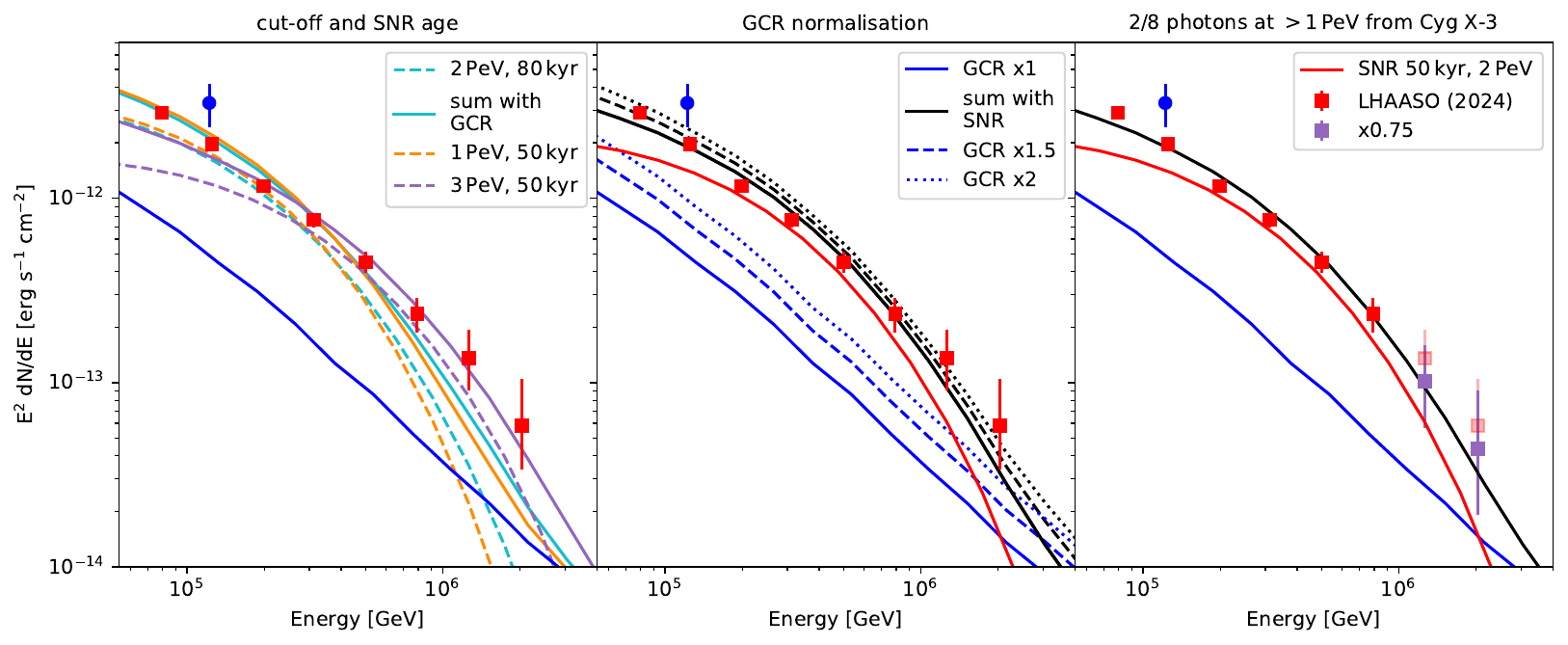}
    \caption{Dependence of the model spectrum on cut-off energy and age of the supernova remnant (left panel) and uncertainty in flux of galactic cosmic rays (middle panel). The right-hand panel shows the spectrum for the case that two of the eight photons with energy ${>}1\,$PeV can be attributed to the microquasar Cygnus~X-3. In addition to a change in spectral shape, the energy required to match the normalisation of the data increases for lower cut-off energies and older supernova remnants. The displayed models require 4.5, 6, and $2.1\times10^{50}\,$erg in non-thermal protons for an age of 80\,kyr, a cut-off energy of 1\,PeV, 3\,PeV, respectively.}
    \label{fig:sed_vary_pars}
\end{figure*}


In Sect.~\ref{sec:SNRsimulation} we concluded that, despite the absence of a detectable remnant, a supernova likely occurred in \ob in the past 10--100\,kyr and inferred the maximum energy of accelerated particles from a hydrodynamic simulation. Based on this estimate (see Fig.~\ref{fig:SNR_histograms}), we assume a maximum energy of 2\,PeV for the following calculation. To model the $\gamma$-ray spectrum of protons accelerated at the forward shock of the remnant,
we follow the approach outlined in Sect.~\ref{sec:modelling}, assuming an exponential cut-off power-law injection, $Q \propto p^{-s} \exp(-p/p_\ur{max})$, where $p_\ur{max}=2\,\ur{PeV}c^{-1}$, the injection momentum is set to  10\,$\ur{MeV}c^{-1}$ and $s=4$. Particles are impulsively injected over a sphere of radius 20\,pc around the centre of \ob.

Figure~\ref{fig:sed_gamma} shows the region-integrated spectra for a supernova remnant age of 50\,kyr and Fig.~\ref{fig:map_gamma} the particle flux map, smoothed using a Gaussian kernel of width $\sigma=0.3^\circ$ to compare to the data. To reproduce the morphology, we place clouds F--K at a distance of 137\,pc behind \ob and the North Filament (cloud E) 140\,pc in front of \ob. For this distribution of target material, the model requires a total energy of $3\times 10^{50}\,$erg in non-thermal protons above 0.01\,GeV. At $100\,$TeV--PeV, the spectrum is steep with slight curvature. In the TeV range, the bulk of the emission comes from interactions with gas inside the superbubble. At higher energies, contributions from both molecular clouds and neutral gas become comparable. The molecular cloud component shows a low-energy cut-off, which stems from the fact that no clouds are located close to the particle injection site and that the diffusion length of protons at these energies is significantly smaller than the distance to the nearest cloud. 

Overall, the observed morphology with a bright spot in the south and an additional weaker bright spot in the north is well reproduced (see Fig.~\ref{fig:map_gamma}). In addition, the model reproduces the energy dependent morphology between the 25--100\,TeV and ${>}100\,$TeV LHAASO bands. The morphology of the emission in the upper and middle LHAASO bands is dominated by emission from molecular clouds. The bulk of the emission comes from the North Filament and Central Clumps in the north. The south is dominated by the SFR South, with some contribution from the Cygnus Far South and Cygnus West clouds. The neutral gas component is required to reproduce the extent and steepness of the radial profile of the emission and the diffuse halo surrounding the clouds (see appendix~\ref{app:suppl_maps}). The energy dependence in the morphology stems the proton distribution, 
which, as shown in Fig.~\ref{fig:fSN}, is more peaked at small radii for 100\,TeV particles than it is for 1\,PeV particles. Consequentially,
the brightness of the neutral gas emission varies with energy. The brightness change is less in molecular cloud emission, since the clouds are located close to the superbubble edge.

\subsection{Supernova remnant age, energy, and target material} 
\label{sec:disc-clouds}

The energy dependence of the spatial profile of the proton distribution function is greater for younger supernova remnants. Ages ${>}80\,$kyr produce minimal energy dependence in the $\gamma$-ray morphology between the 25--100\,TeV and ${>}100\,$TeV bands (see Fig.~\ref{fig:maps_80kyr} in the appendix). We note that the parameters $t_\ur{SN}$ and $D_\ur{SB}$ are degenerate as the proton distribution scales with $\sqrt{D_\ur{SB}t_\ur{SN}}$. A different assumption on $D_\ur{SB}$ will therefore affect the age estimate.     

The required energy corresponds to a powerful supernova with $3\times10^{51}\,$erg, if a proton acceleration efficiency of 10\% is assumed. A powerful supernova is consistent with the assumption made in Sect.~\ref{sec:SNRsimulation} to motivate a cut-off energy of a few PeV. The energy requirement can be reduced by assuming a superbubble density ${>}0.05\pccm$, a younger supernova remnant, or, in principle, a higher cut-off energy. For younger (more recent) supernova remnants, particle escape at high energies is reduced, which leads to higher normalisations, harder spectra, and a cut-off at higher energy. 
A lower bound for the supernova age is probably 10--20\,kyr, given that no remnant is detected in \ob (see Sect.~\ref{sec:sn-evidence}).

In the scenario of higher superbubble density, molecular clouds would have to be slightly further embedded in the superbubble to maintain the relative normalisation between the neutral gas and molecular cloud component, which is constrained by the $\gamma$-ray morphology (see Fig.~\ref{fig:maps_suppl} in the appendix). For typical superbubble densities, $n_\ur{SB}\lesssim0.1\pccm$, the model predicts the clouds to be within ${\sim}10\mbox{--}20\,$pc of the edge. We note that the normalisation of the proton distribution changes rapidly across the superbubble edge (see Fig.~\ref{fig:fSN}). This is a consequence of the inhomogeneous two-zone diffusion model. A change in cloud position within a few parsecs can therefore produce measurable differences in the $\gamma$-ray maps. We highlight however that in fitting the cloud positions, we only deviated from the assumption of equiradial clouds in the case of the North Filament, which is located at a radius of 140\,pc instead of 137\,pc as is the case for the remaining clouds in the vicinity of \ob. In other words, except for the North Filament, the $\gamma$-ray brightness of clouds in our model is well correlated with the cloud mass. We remark in addition that cloud masses are subject to uncertainty (see Sect.~\ref{sec:clouds}). As highlighted in Sect.~\ref{sec:modelling}, approximately equiradial clouds close to the superbubble edge are expected. The fact that this assumption holds well within our model supports the validity of the supernova scenario.

\subsection{Origin of the ultra-high energy $\gamma$-rays} 
\label{sec:uhe}

Substantial interest has been taken in the origin of the ultra-high energy $\gamma$-rays from the Cygnus region. In the following, we explore how well the supernova remnant model describes the data up to the highest energies. Figure~\ref{fig:sed_vary_pars} highlights relevant uncertainties in the modelling: age of the supernova remnant and cut-off energy, the normalisation of the galactic cosmic ray background, and the contribution of Cygnus X-3 to the flux above 1\,PeV. Our default assumption of $E_\ur{max}=2\,$PeV and $t_\ur{SN}=50$\,kyr gives a model within the $1\sigma$ uncertainty of the two highest-energy data points. Models with a lower cut-off or older supernova remnants miss these points (see Fig.~\ref{fig:sed_vary_pars}, left panel). 
Our choice of $E_\ur{max}=2\,$PeV is motivated for a powerful supernova (cf.~Fig.~\ref{fig:SNR_histograms}).
For a nominal supernova, a lower cut-off energy might be more appropriate (about 0.5-1\,PeV for an explosion energy of $10^{51}$\,erg assuming $E_\ur{max}$ is roughly proportional to the square root of the explosion energy). However, as argued in Sec.~\ref{sec:sn-evidence}, the young age of Cygnus~OB2 favours Type Ic supernovae whose energy likely exceeds $10^{51}$\,erg \citep{Fryer_2018}.

The background from diffuse galactic cosmic rays and the possibility of PeV particle acceleration in Cygnus~X-3 introduce further uncertainty into the model. The uncertainty in the galactic cosmic-ray component is about a factor of two \citep{Schwefer2023}. With an increase in the background within this range, the constraints on supernova remnant age and cut-off energy can be relaxed slightly. However, models with high galactic cosmic-ray background tend to over-predicted the flux at TeV energies if the \fmlat data is fit with inverse Compton emission from stellar winds in \ob (see Sect.~\ref{sec:gamma-ob2}) and require enhanced synchrotron cooling, $B\gtrsim10\,$\textmu G. 

Microquasars have recently received increased attention as potential accelerators of PeV particles. While at a larger distance, Cygnus X-3 is in spatial coincidence with \ob and two out of the eight photons reported in \citet{Lhaaso2024} above 1\,PeV are spatially correlated with its location. In the right panel of Fig.~\ref{fig:sed_vary_pars}, we naively scale the flux points above 1\,PeV by a factor 2/8, leaving the error bars unchanged. This is meant only to illustrate how attributing some of the detected photons to Cygnus X-3 might affect the flux expected from the supernova remnant and should not be seen as proper modelling. A cut-off of 2\,PeV provides the best description of the scaled data points (unscaled: 3\,PeV). We note, however, that the uncertainties are large and that the flux from diffuse galactic cosmic rays is almost within $1\sigma$.    

In conclusion, considering that an association of at least some of the ultra-high energy photons with Cygnus~X-3 is likely and that the significance at high energies is low, the supernova remnant scenario is conceivable even for a lower cut-off energy. With standard assumptions (no contribution from Cygnus~X-3 and standard galactic cosmic-ray background) a cut-off energy of 2\,PeV is required to remain within the $1\sigma$ uncertainties of the data above 1\,PeV, if the injection spectrum is a $p^{-4}$ power-law with exponential cut-off.

\subsection{Sensitivity to other parameters} 

We set $D_{\rm ISM} = 10^{28} p_{\rm GeV}^{0.4}\ur{cm}^2\,\ur{s}^{-1}$ and the superbubble radius to 150\,pc. A change in these parameters within realistic bounds only affects the model marginally. Key to the behaviour of the model is that $D_\ur{ISM} \gg D_\ur{SB}$, which remains true even if $D_\ur{ISM}$ is reduced, for example, by a factor ten. Considering the discussion in Sect.~\ref{sec:target-gas}, the superbubble radius might vary by as much as $20\mbox{--}30\%$, depending on properties of the environment and star cluster. A change of this magnitude does not affect the shape of the particle distribution significantly.

Finally, in our model we fix the spectral index at injection to the nominal value $s=4$ obtained from test-particle diffusive shock acceleration theory \cite[e.g.][]{drury1983}. This parameter is degenerate with other model parameters, in particular the supernova age, the positions of molecular clouds relative to the superbubble edge, and the diffusion coefficient in the superbubble. If other model parameters are kept fixed, steepening, be it from injection or propagation effects, reduces the fit quality at ultra-high energies. As discussed in Sect.~\ref{sec:uhe}, this can be compensated for by contributions from Cygnus~X-3 or diffuse galactic cosmic rays.

\subsection{Leptonic component}

\begin{figure*}
    \centering
    \includegraphics[width=\linewidth]{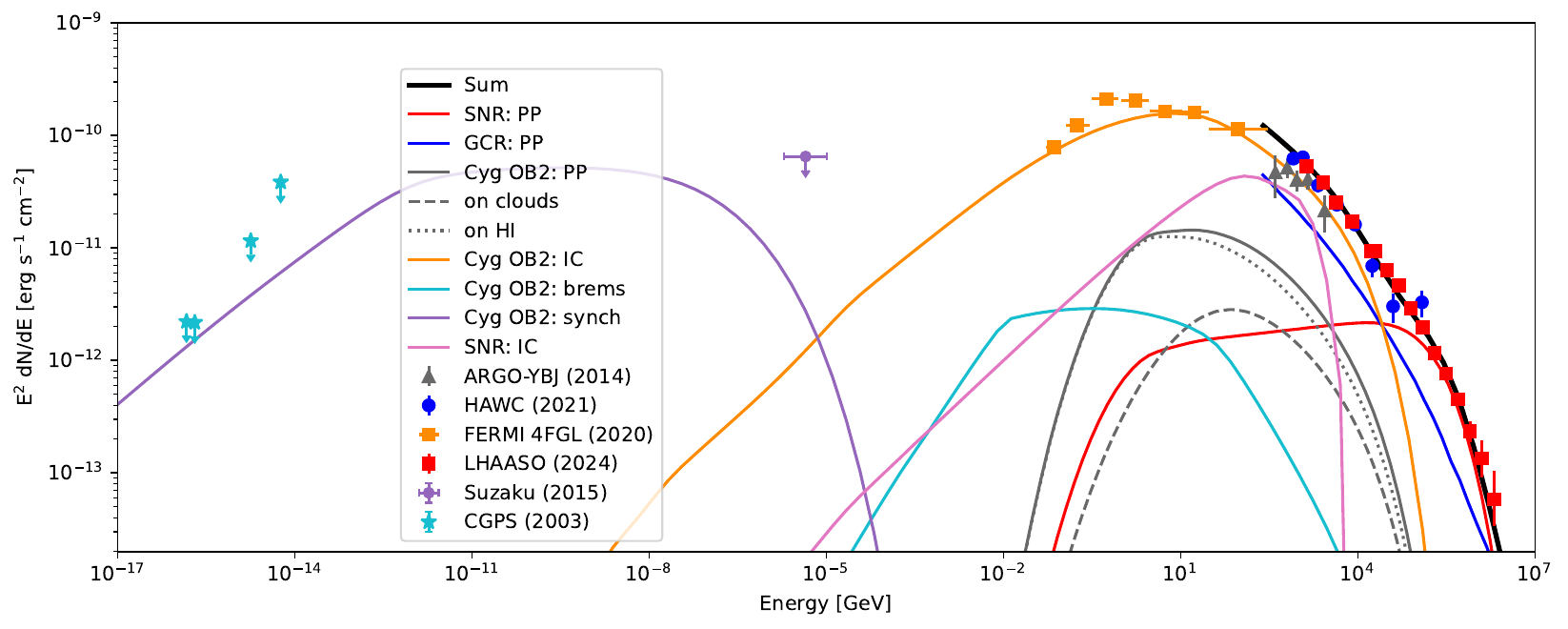}
    \caption{Spectral energy distribution model, showing additional components: synchrotron emission (purple) and bremsstrahlung (cyan) from electrons accelerated at stellar-wind termination shocks in \ob, hadronic emission from the same source (grey), and inverse Compton emission from the supernova remnant shock (pink). The latter is normalised assuming a supernova energy of $3\times10^{51}\,$erg and an acceleration efficiency of electrons of 1\%. We show radio upper limits from \citet{gammaCygnus_HAWC2021}, which were derived for the entirety of the Cygnus Cocoon. We assume uniform $B=5\,$\textmu G and $n=0.05\pccm$ in the superbubble.}
    \label{fig:sed_broad}
\end{figure*}

Electrons accelerated by the supernova remnant shock will produce inverse-Compton emission. This results in a component with a comparatively sharp peak at about 1\,TeV, which cuts off at higher energies due to synchrotron losses, as shown in pink in Fig.~\ref{fig:sed_broad} for an electron acceleration efficiency of 1\% and assuming the same photon fields as described in Sect.~\ref{sec:phfields}. Leptonic emission from the supernova is subdominant compared to the much broader underlying inverse Compton component from stellar winds, such that a discernable observational signature in $\gamma$-rays is not expected.

\section{Leptonic emission from \ob}
\label{sec:gamma-ob2}

\subsection{Results and discussion}

\ob harbours several early-type and Wolf-Rayet stars, whose stellar-wind termination shocks are expected to accelerate particles up to energies of the order of 50\,TeV (see Sect.~\ref{sec:candidate_accelerators}). We obtain the particle distribution and $\gamma$-ray spectrum as described in Sect.~\ref{sec:modelling}, taking into account the radial dependence of the cluster photon field for computing losses and $\gamma$-ray emission. Particles are injected continuously over 3\,Myr within a sphere of radius 10\,pc centred on \ob. 
The injection spectrum is a power-law with an exponential cut-off at $p_\ur{max}=50\,\ur{TeV}c^{-1}$ and a spectral index $s=4.1$. The injection momentum is set to 10\,$\ur{MeV}c^{-1}$, as in the supernova model. The choice for the spectral index is motivated by the observation of comparatively steep $\gamma$-ray spectra from star-forming environments \citep[e.g.][]{hess-wd1-21, hess-lmc-24, lhaaso24-w43}. Figure~\ref{fig:sed_gamma} shows the stellar-wind inverse Compton component in yellow. The model reproduces the hardening of the spectrum in the \fmlat band and the softening above 100\,GeV. The $\gamma$-ray morphology is expected to peak at \ob and extend to a radius corresponding to the distance travelled by electrons within their cooling time (see Fig.~\ref{fig:ic-losses}), which is ${\sim}50\,$pc in our model. This is consistent with the morphology observed by \citet{Lhaaso2024} in the 2--20\,TeV band.

To reproduce the flux level observed in $\gamma$-rays, the required power in non-thermal electrons is $3.5 \times 10^{36}\ergs$ across all energies, which represents about 1\% of the currently inferred stellar wind power in Cygnus~OB2. While this value is higher than that typically inferred from multi-wavelength models of supernova remnants \cite[e.g.][]{Berezhko}, the shock conditions at termination shocks of isolated stellar winds are quite different to those at the external shocks of young supernova remnants. 
For example, they have a rather slow upstream flow speed and high upstream temperature compared to supernova remnants. 
Stellar wind termination shocks are also expected to show high magnetic field obliquity.
Furthermore, since electrons are injected continuously over the lifetime of the association, the expected flux depends on past stellar wind activity. It is conceivable that the wind power of \ob was higher in the past than the currently inferred $3\times10^{38}\ergs$, considering the complex star-formation history of the association (see Sect.~\ref{sec:ob2}).

The bottom panel of Fig.~\ref{fig:sed_gamma} shows the contributions from scattering on different photon fields. In the \fmlat band, the scattering on the cluster photon field dominates. The TeV range is dominated by scattering on dust emission. Figure~\ref{fig:sed_broad} shows the multi-wavelength spectrum, including radio upper limits for the Cygnus Cocoon \citep{gammaCygnus_HAWC2021} and X-ray upper limits from \textit{Suzaku}. For our assumption of a homogeneous magnetic field of 5\,\textmu G, the synchrotron model remains within the upper limits, while bremsstrahlung is always subdominant in the low-density superbubble.

Our choice of parameters provides a good fit to the LHAASO and \fmlat data in the ${\sim}1$\,GeV--10\,TeV range while remaining within the multi-wavelength limits. Considerable uncertainty stems from the fact that the magnetic field is not well constrained and most certainly varies within the region. Due to the non-trivial nature of stellar wind interaction \citep{Vieu2024CygnusSimu, Haerer2025}, it is challenging to make an informed choice for a prescription for the spatial dependence of the magnetic field. It is conceivable that the magnetic field in the inverse Compton emission region (radii ${\lesssim}50\,$pc) is larger than 5\,\textmu G but decreases beyond to result in the given radio upper-limits. Increased synchrotron losses soften the spectrum above 1\,TeV. Best-fit models for $B\gtrsim 10\,$\textmu G require either harder injection spectra ($s=4$) or increased dust emission. As mentioned we already assume that dust emission is enhanced by a factor of two over the model by \citet{popescu17} as is appropriate in spiral arms. It is however conceivable that the enhancement factor is higher in the vicinity of \ob due to local dust production in the star-forming environment \citep[see][]{green19}. Models with $B<5\,$\textmu G provide a good fit for softer particle injection spectra, but tend to over-predict the $\gamma$-ray flux at 100\,MeV--1\,GeV. Alternatively, the superbubble flow velocity could have a profile shallower than $r^{-2}$, in which case adiabatic losses would reduce the flux back within range. This is one of several factors that limits the predictive power of the model at low energies. The model also depends on the choice of lowest particle momentum (10\,MeV c$^{-1}$ here) and the particle injection history over several millions of years. \ob harbours several distinct stellar populations, with ages in the range of 3--7\,Myr \citep[see][]{Vieu2024CygnusSimu}. Particle injection was almost certainly not continuous. 
Jointly, these factors can lead to a more complex spectral shape than predicted by our model. The fact that the bump around 1\,GeV is not captured by the model is therefore not surprising. We note that assumptions on the low-energy electron population also impact synchrotron emission in the radio.

\subsection{Hadronic component}
\label{sec:gamma-additional}

The grey lines in Fig.~\ref{fig:sed_broad} show $\gamma$-rays produced in molecular clouds and in the neutral medium by protons accelerated at stellar-wind termination shocks in \ob. We assume a hadronic acceleration efficiency of 10\%. The model is dominated by the flux from hadronic interactions within the superbubble interior, 
because particles diffuse away quickly once they leave the superbubble and enter the interstellar medium. Therefore, they only produce low levels of $\gamma$-ray flux within the higher-density regions beyond the cavity.
Simultaneously, the continuous supply of fresh particles from \ob ensures that the density of particles inside the superbubble remains high.
The model is below the measured flux by about one order of magnitude. A fit to the data would require an unusually high superbubble interior density, $n_\ur{SB}>0.5\pccm$. For instance, previous claims that Cygnus~OB2 produces hadronic $\gamma$-rays matching the observed flux assume densities of $20\mbox{--}30$\,cm$^{-3}$ \citep[][]{gammaCygnus_HAWC2021,Bykov2022Cygnus,Banik2022,Menchiari2024}. 
Within the framework of \citet{weaver1977} $n_\ur{SB}>1\pccm$ is only obtained assuming Cygnus~OB2 was born in a parent molecular cloud of density ${>} 10^3$\,cm$^{-3}$. This is two orders of magnitude larger than the mean value estimated from the HI column density.

\section{Conclusion}
\label{sec:conclusion}

We demonstrated that a powerful supernova, which exploded about 50\,kyr ago in the stellar association \ob is the best candidate 
for the origin of the diffuse $\gamma$-ray emission at ${\sim}10$\,TeV--PeV observed in the Cygnus region. Our model reproduces the observed energy dependence of the $\gamma$-ray morphology. With an additional inverse Compton component from electrons accelerated at stellar-wind termination shocks in \ob, we model the full $\gamma$-ray spectrum from 100\,MeV--PeV. This joint lepto-hadronic scenario is based on the following considerations:
\begin{enumerate}
    \item Cygnus is a vast, close-by star-forming complex. Multi-wavelength observations reveal large-scale structures shaped by feedback from generations of massive stars. The main source of mechanical power is the loose stellar association \ob. Even though the connection between large-scale structures observed at different wavelengths is far from being understood, they support the notion of a low-density superbubble about 150\,pc in radius around \ob. 
    \item Cygnus harbours powerful objects, such as Wolf-Rayet stars, pulsars, and supernova remnants. None of the detected objects stand out as a clear candidate source of ultra-high energy $\gamma$-ray emission. We argue that an undetected supernova with above-average power has likely occurred within the last few 100\,kyr.
    Extending the 3D hydrodynamical simulations presented in \citet{Vieu2024CygnusSimu}, we determine the maximum particle energy at the forward shock of such a supernova remnant to be 1--2\,PeV. 
    \item We consider a two-zone diffusion model, where the first zone represents the highly turbulent superbubble and the second zone the interstellar medium.   
    A low-diffusion zone is required to confine the non-thermal particles within the $\gamma$-ray emission region. We obtain the particle distribution function by numerically solving the transport equation in spherical symmetry. Reviewing the literature, we identify six molecular clouds within 100--200\,pc from \ob, which serve as target material for $\gamma$-ray production by non-thermal protons. In our preferred model, these clouds are located about equiradially at the edge of the \ob superbubble. 
\end{enumerate}
In the proposed scenario, an old supernova remnant would remain at the present day, which, reaching a size of several 10s of parsecs in a hot, low-density superbubble may have escaped detection.
Nevertheless, its past feedback would not only impact the production of very-high energy cosmic rays, but also have implications for our understanding of the dynamics of star-forming complexes. In the case of Cygnus, this might be key to understanding how the complex was carved up to large-scales.
As mentioned in Sec.~\ref{sec:candidate_accelerators}, it is also not excluded that the very-high energy particles were accelerated during the early expansion phase of the $\gamma$-Cygni supernova remnant which, despite its offset, might be embedded in the Cygnus~OB2 superbubble. Very-high energy particles could have leaked out of the remnant, filled the superbubble, and reached the complex of molecular clouds in Cygnus~X. 
Determining under which conditions $\gamma$-Cygni can contribute to the non-thermal content of the Cygnus~OB2 superbubble at very-high energy
requires a detailed, asymmetric, propagation model, which 
is beyond the scope of the present work.

While a powerful supernova can account for the $\gamma$-ray photons observed at ultra-high energies, an association of all flux above 1\,PeV with such an event is not the only plausible scenario. Some ultra-high energy photons are spatially correlated with the microquasar Cygnus~X-3 and the contribution from diffuse galactic cosmic rays at ultra-high energies is uncertain. Taking these factors into account allows for a cut-off in the supernova remnant $\gamma$-ray spectrum below 2\,PeV, in other words, the existence of an extreme accelerator in Cygnus~X is not essential to explain the observations. 
At hundreds of TeV, the supernova remnant scenario is preferred over the wind termination shock scenario, since \ob is too extended to form a collective cluster wind and the maximum particle energy at individual stellar-wind termination shocks is at most 100\,TeV. 
Particles accelerated at wind shocks are however the preferred source of $\gamma$-rays below ${\sim}10$\,TeV. Because of the continuous supply of freshly accelerated particles, strong photon fields, and the low-density superbubble environment, inverse Compton dominates over hadronic emission and allows to explain the data with an electron acceleration efficiency of 1\%.

Given the current statistical uncertainties, future observations with LHAASO and CTAO are necessary to constrain the origin of ultra-high energy $\gamma$-ray photons observed in Cygnus. In addition, a dedicated multi-wavelength effort is needed to decipher the connection between large-scale structures across the electromagnetic spectrum in the Cygnus region. 
The relationship between measurements in the ultra-high energy $\gamma$-ray regime and the role of massive-star forming regions in the galactic cosmic-ray ecosystem remains an open problem. In this work, we established a connection between an energetic supernova in \ob and multi-PeV cosmic rays.

{\bigbreak \tiny \noindent
\textit{Data availability.}
The column density maps for the ten molecular clouds listed in Table~\ref{tab:MolecularClouds} are available at \url{https://doi.org/10.17617/3.G9LUS1}. They were inferred using the CO data and the catalogue of cloud distances from \citet{Zhang2024} and are available at \url{https://doi.org/10.57760/sciencedb.16716} and \url{https://www.doi.org/10.26093/cds/vizier.51670220}, respectively.
The initial conditions and output of the hydrodynamic simulation (Sect.~\ref{sec:simu}) may be shared on
reasonable request to TV.
Details on the transport equation solver (Sect.~\ref{sec:prop}) may be shared on reasonable request to FS (florian.schulze@mpi-hd.mpg.de) or TV.
\medbreak
}

\begin{acknowledgements}
      This work made use of the MHD code PLUTO. Computations were performed on the HPC system Raven at the Max Planck Computing and Data Facility. We thank L.~Olivera Nieto and Q.~Remy for numerous helpful discussions on the $\gamma$-ray data and the gas distribution. We acknowledge R.~Yang for providing LHAASO significance maps and details on the LHAASO analysis. CJKL gratefully acknowledges support from the International Max Planck Research School for Astronomy and Cosmic Physics at the University of Heidelberg in the form of an IMPRS PhD fellowship. 
\end{acknowledgements}

%
%

\bibliographystyle{aa}
\bibliography{biblio}


\begin{appendix}
\onecolumn

\section{Distance to Cygnus~OB2}
\label{appendix:distancetoOB2}

Cygnus~OB2 is by far the most massive and powerful association of massive stars in the Cygnus region. The determination of its distance is critical to understanding feedback processes on nearby molecular clouds and surrounding multi-wavelength emission, from radio to ultra-high energy $\gamma$-rays. Spectrophotometric studies usually show a large spread in the distance modulus of the association and do not allow positioning of the association accurately within the Cygnus-X region \citep[e.g.][]{Hanson2003}.

\begin{figure*}[h]
	\sidecaption
	\centering
	\includegraphics[width=0.7\linewidth]{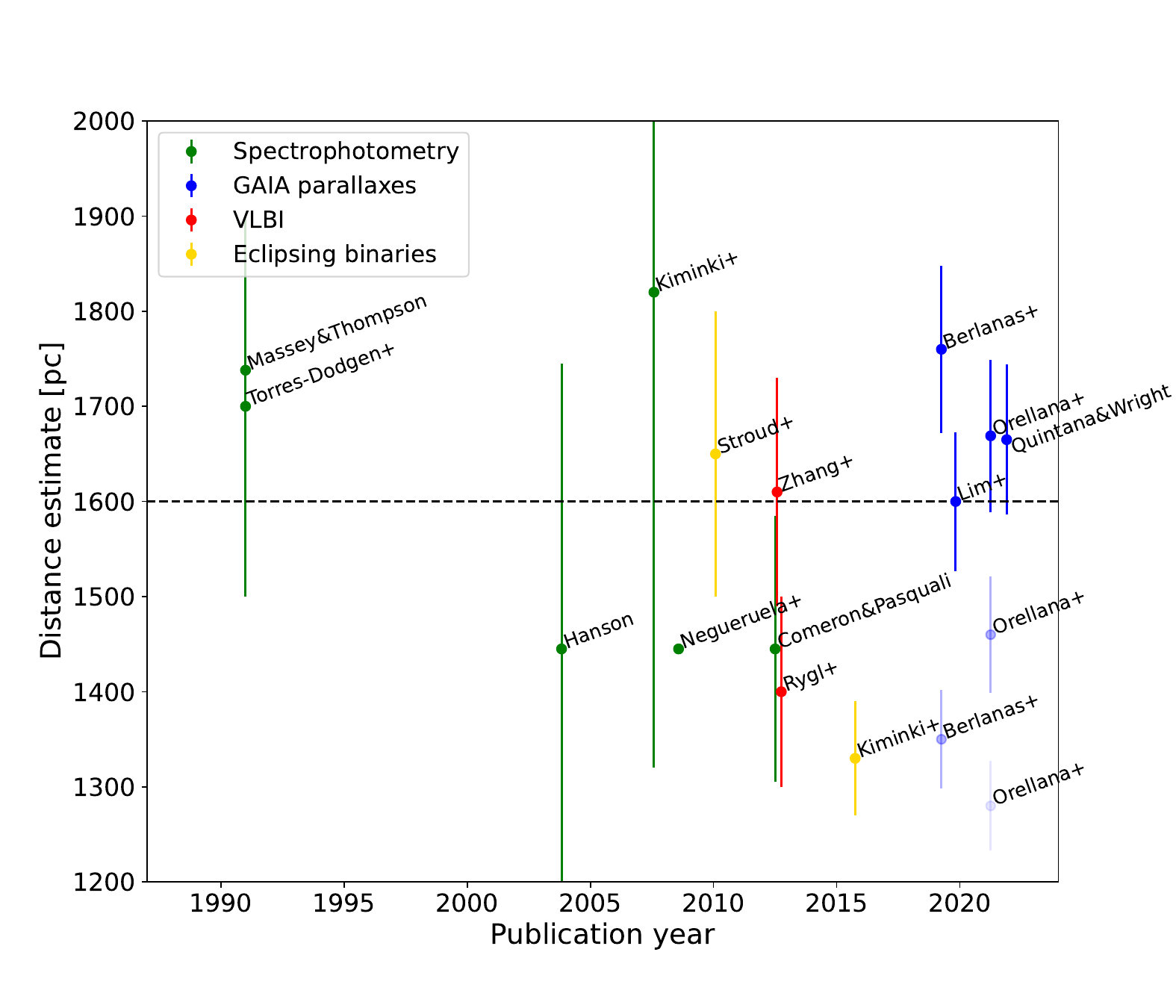}
	\caption{\label{fig:DistanceOB2}Distance estimates for Cygnus OB2 quoted in the literature. Distances to foreground substructures are shown in fading colours. A systematic uncertainty of 0.03\,mas is added to quoted {\textit{Gaia}} parallaxes.}
\end{figure*}

Figure~\ref{fig:DistanceOB2} compiles distance estimates found in the literature. Very Long Baseline Interferometry (VLBI) measurements of masers in Cygnus North suggest a distance of $1.3\mbox{--}1.5$\,kpc for these molecular clouds \citep{Rygl2012}, however this might not necessarily be the distance to Cygnus~OB2. VLBI measurements of circumstellar masers around the red hypergiant NML~Cyg provide a distance of $1.61 \pm 0.12$\,kpc, but NML~Cyg, with a latitude of -1.9$^\circ$, does not exactly correlate with Cygnus~OB2; its projected separation is about 90\,pc.

\citet{Kiminki2015} measure the distance to four spectroscopic binaries at $1.33 \pm 0.6$\,kpc. However, {\textit{Gaia}}~EDR3 provides compatible parallaxes for three of these binaries, which are consequently no longer considered to be in the main group of Cygnus~OB2. Only for one binary, MT91~372, there is tension in the distance determination ($1765 \pm 44$\,pc according to {\textit{Gaia}} EDR3, still placing it within Cygnus~OB2). A dedicated study of the eclipsing blue-supergiant binary V$^\star$~V1827~Cyg finds a distance of $1.65 \pm 0.15$\,kpc, in good agreement with the {\textit{Gaia}} EDR3 parallax measurement ($1765 \pm 70$\,pc). This star is not close to the centre of Cygnus~OB2, but at a projected distance of 0.45$^\circ$ (13\,pc at a distance of 1.7\,kpc). Still, it is considered to be part of the association in modern censuses.

Despite the systematic bias, {\textit{Gaia}} parallaxes seem to provide the most direct, robust and accurate distance measurements. The latest analyses \citep{Quintana2021,Orellana2021} are compatible with distances of $1.6\mbox{--}1.75$\,kpc. Three stellar foreground groups are identified by \citet{Orellana2021}, one at around 1.46\,kpc, which is associated with the open cluster UBC~585, and two at around 1.28\,kpc.

\section{Supplementary $\gamma$-ray maps}
\label{app:suppl_maps}

\begin{figure*}
    \centering
    \includegraphics[height=0.38\linewidth]{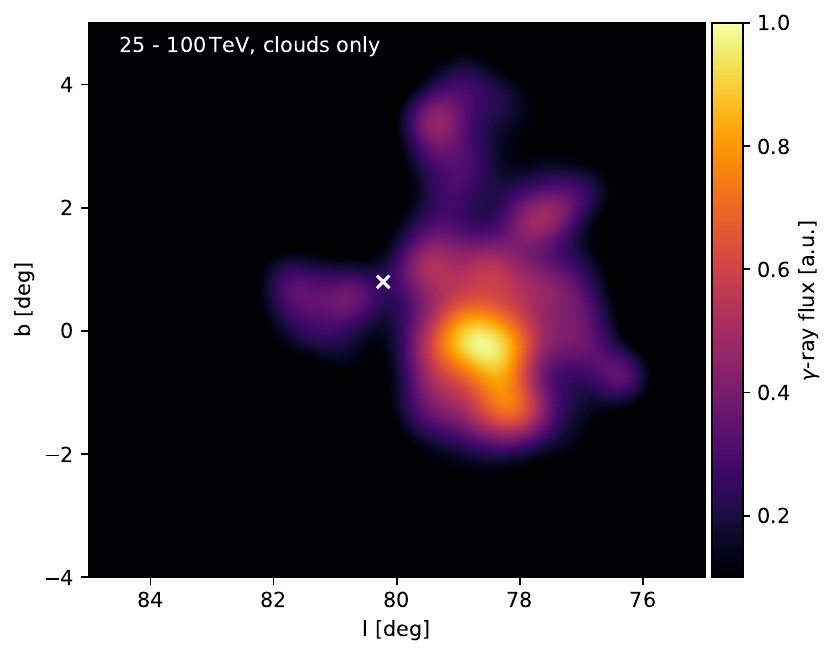}
    \includegraphics[height=0.38\linewidth]{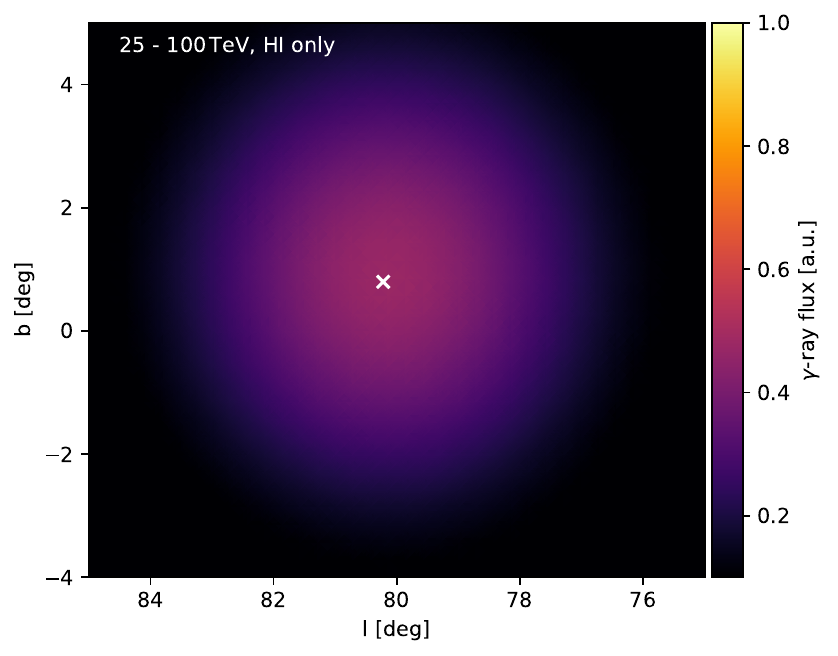}
    \includegraphics[height=0.38\linewidth]{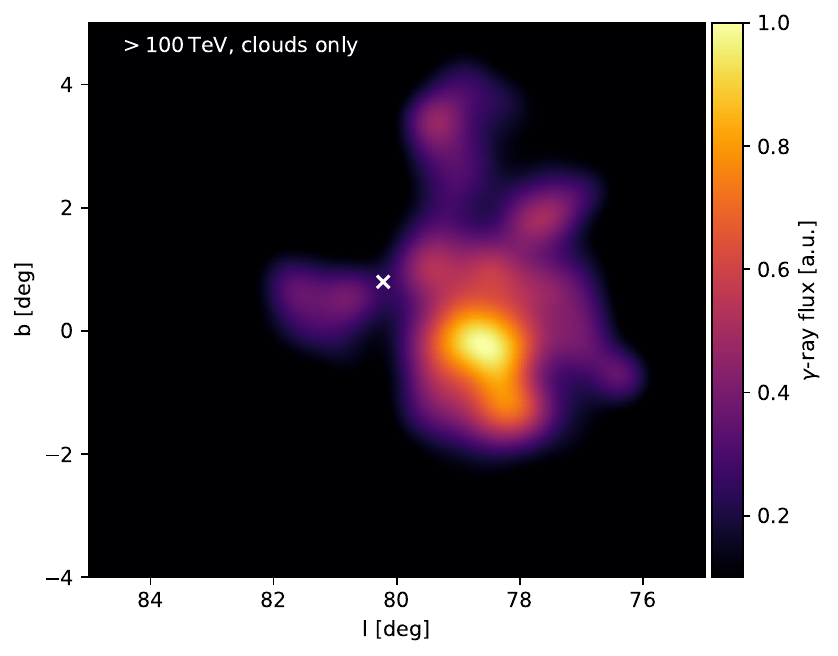}
    \includegraphics[height=0.38\linewidth]{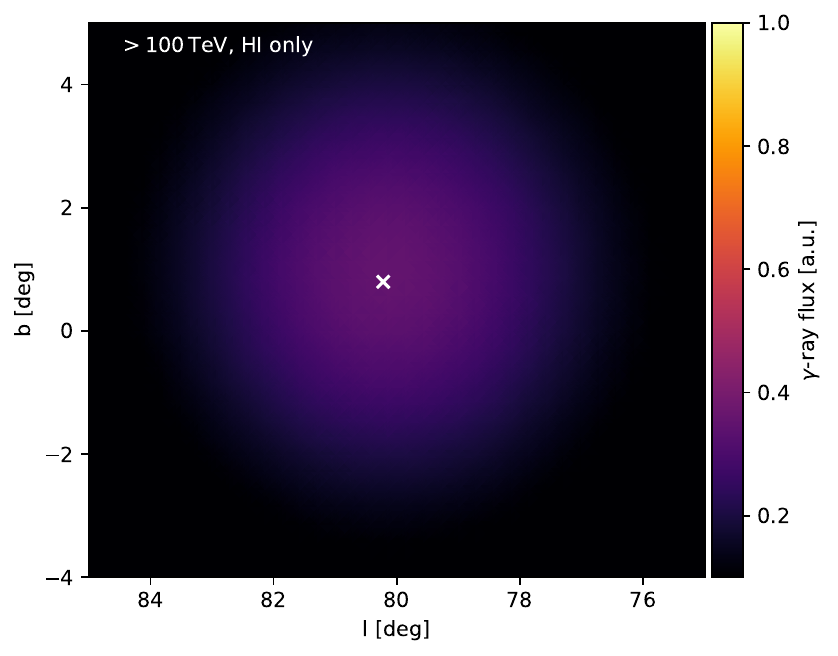}
    \caption{Components of the synthetic $\gamma$-ray emission maps shown in Fig.~\ref{fig:map_gamma}. 
    The colour-scale is individually normalised to the maximum in each energy band. Maps are smoothed using a Gaussian function with a half-width of 0.3$^\circ$.}
    \label{fig:maps_suppl}
\end{figure*}

\begin{figure*}
    \centering
    \includegraphics[height=0.38\linewidth]{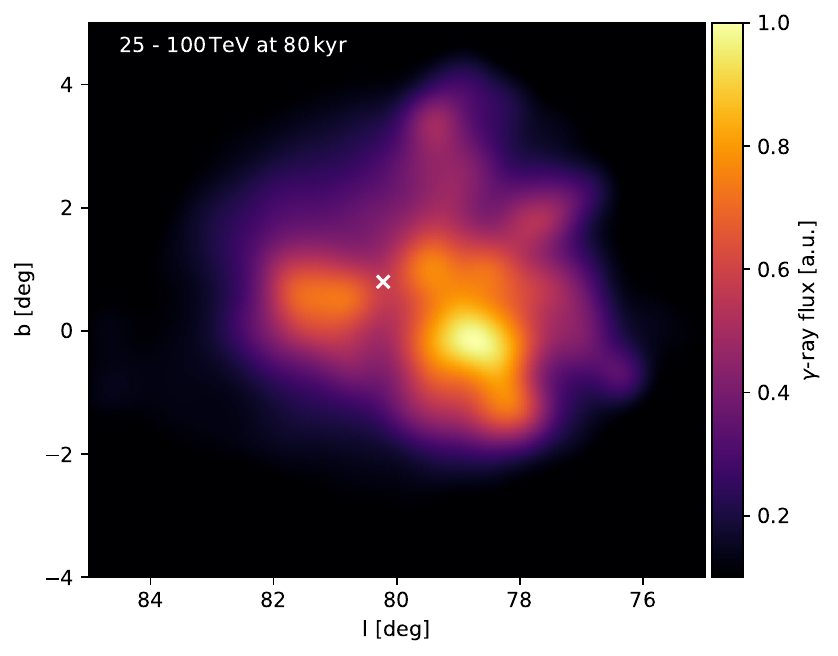}
    \includegraphics[height=0.38\linewidth]{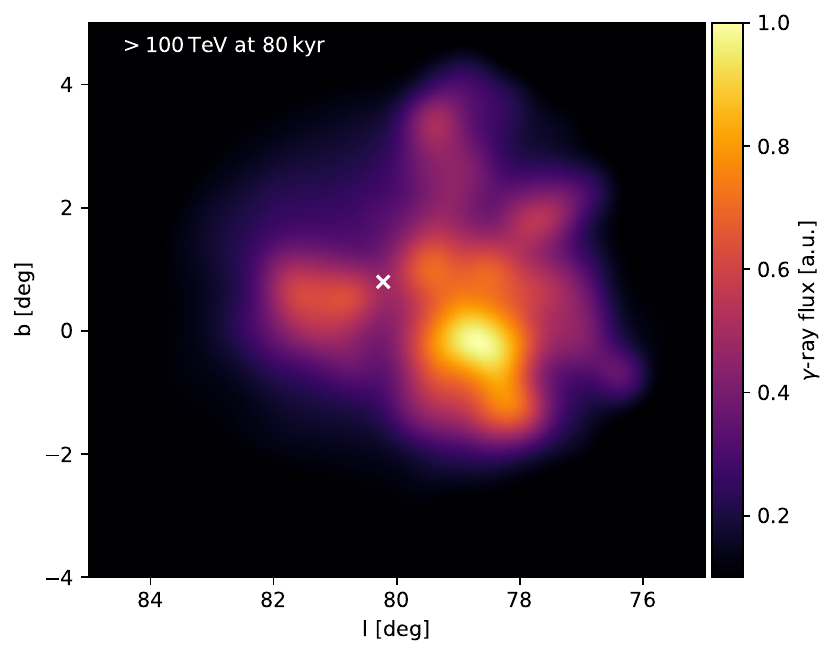}
    \caption{Model at 80\,kyr, which shows that older supernova remnants produce less prominent energy dependence in the northern clouds at $l>80^\circ$ (cf.~Fig.~\ref{fig:map_gamma}). The colour-scale is individually normalised to the maximum in each energy band. Maps are smoothed using a Gaussian function with a half-width of 0.3$^\circ$.}
    \label{fig:maps_80kyr}
\end{figure*}


Figures~\ref{fig:maps_suppl} and \ref{fig:maps_80kyr} show the synthetic $\gamma$-ray emission for the supernova remnant model (see Sect.~\ref{sec:gamma-sn}). Figure~\ref{fig:maps_suppl} shows components of the model presented in Fig.~\ref{fig:map_gamma}. The figure demonstrates that the diffuse neutral gas component (HI) is required to reproduce the morphology of the emission, since emission from the clouds alone is too localised. In addition, the figure highlights that the HI component is brighter at lower energies. This is discussed in Sect.~\ref{sec:gamma-sn} and also shown in Fig.~\ref{fig:fSN}. Figure~\ref{fig:maps_80kyr} shows the model at 80\,kyr. While the morphology in the ${>}100\,$TeV band is unchanged compared to the model with $t_\ur{SN}=50\,$kyr, the 25--100\,TeV band shows decreased brightness in the northern clouds at $l>80^\circ$. As a consequence, the northern clouds are only marginally brighter in the lower-energy band than in the higher-energy band and match the LHAASO observations less well (see Fig.~\ref{fig:map_gamma}).  

\end{appendix}
\end{document}